\documentclass[prd,twocolumn,amsmath,amssymb,floatfix,superscriptaddress,nofootinbib,preprintnumbers]
{revtex4-1}
\usepackage[utf8]{inputenc}
\usepackage{amssymb}
\usepackage{amsmath}
\usepackage[toc,page]{appendix} 
\usepackage{bm}
\usepackage{color}
\usepackage{enumitem}
\usepackage{booktabs}
\usepackage{subcaption}
\usepackage{graphicx}
\usepackage[labelfont=bf]{caption}
\usepackage{pifont}
\usepackage[table, svgnames, dvipsnames]{xcolor}
\usepackage{makecell, cellspace, caption}
\definecolor{aliceblue}{rgb}{0.94, 0.97, 1.0}
\definecolor{cambridgeblue}{rgb}{0.64, 0.76, 0.68}
\definecolor{celadon}{rgb}{0.67, 0.88, 0.69}
\definecolor{classicrose}{rgb}{0.98, 0.8, 0.91}

\DeclareMathAlphabet\mathbfcal{OMS}{cmsy}{b}{n}
\definecolor{darkgreen}{RGB}{50,150,0}
\definecolor{purple}{cmyk}{0.5,0.75,0,0}
\definecolor{darkpurple}{RGB}{128,0,128}
\definecolor{ultramarine}{rgb}{0.07, 0.04, 0.56}
\definecolor{cadmiumgreen}{rgb}{0.0, 0.42, 0.24}
\definecolor{indigo(dye)}{rgb}{0.0, 0.25, 0.42}

\usepackage[linktocpage=true]{hyperref}
\hypersetup{
colorlinks=true,
citecolor=ultramarine,
linkcolor=cadmiumgreen,
urlcolor=indigo(dye),
pdfauthor={},
pdftitle={},
pdfsubject={}
}

\def\apj{{\rm ApJ}}                 % Astrophysical Journal
\def\aap{{\rm A\&A}}                % Astronomy and Astrophysics
\def\mnras{{\rm MNRAS}}             % Monthly Notices of the RAS
\def\pasp{{\rm PASP}}               % Publications of the ASP
\def\mnras{Mon.\ Not.\ R.\ Astron.\ Soc.\ }

\begin{document}
%\preprint{YITP-SB-16-18}

\title{Direct Detection of Dark Matter Substructure in Strong Lens Images with Convolutional Neural Networks}

\author{Ana Diaz Rivero}
\email{adiazrivero@g.harvard.edu}
\affiliation{Department of Physics, Harvard University, Cambridge, MA 02138, USA}

\author{Cora Dvorkin}
\email{cdvorkin@g.harvard.edu}
\affiliation{Department of Physics, Harvard University, Cambridge, MA 02138, USA}

\begin{abstract}
\noindent Strong gravitational lensing is a promising way of uncovering the nature of dark matter, by finding perturbations to images that cannot be well accounted for by modeling the lens galaxy without additional structure, be it subhalos (smaller halos within the smooth lens) or line-of-sight (LOS) halos. We present results attempting to infer the presence of substructure from images without requiring an intermediate step in which a smooth model has to be subtracted, using a simple convolutional neural network (CNN). We find that the network is only able to infer the presence of subhalos with $>75\%$ accuracy when they have masses of $\geq 5 \times 10^9$M$_{\odot}$ if they lie within the main lens galaxy. Since less massive foreground LOS halos can have the same effect as higher mass subhalos, the CNN can probe lower masses in the halo mass function. The accuracy does not improve significantly if we add a population of less massive subhalos. With the expectation of experiments such as HST and Euclid yielding thousands of high-quality strong lensing images in the next years, having a way of analyzing images quickly to identify candidates that merit further analysis to determine individual subhalo properties while preventing extensive resources being used for images that would yield null detections could be very useful. By understanding the sensitivity as a function of substructure mass, non-detections could be combined with the information from images with substructure to constrain the cold dark matter scenario, in particular if the sensitivity can be pushed to lower masses. 
\end{abstract}

\maketitle

\section{Introduction}\label{sec:introduction}

For many years the Lambda Cold Dark Matter ($\Lambda$CDM) paradigm has been lauded for its ability to explain vastly different observations of the universe, from galaxy clustering \cite{BOSS} and supernovae luminosities \cite{Pantheon}, to the shape of the power spectrum of the cosmic microwave background (CMB) \cite{Planck2018}. However, as high-precision measurements from a wide variety of surveys come in, possible inconsistencies are being uncovered, and the $\Lambda$CDM model is being scrutinized as never-before. For example, the $H_0$ and $S_8$ tensions $-$ the values of these parameters inferred from local measurements or weak lensing, respectively, differ from those measured by the CMB at the level of a few sigma, e.g. \cite{S8_tension,Tensions} $-$ are two of the most notorious issues that have brought about the day of reckoning for the standard cosmological model.  

Another feature of the $\Lambda$CDM model that has drawn widespread attention is the distribution of dark matter on sub-galactic scales. Many dark matter theories that are consistent on large scales have very different behaviors on subgalactic scales. However, observing these scales poses several challenges that do not tend to plague large-scale observations to the same extent. Star formation becomes increasingly inefficient as halo mass is decreased, meaning that observing the low-mass end of the halo mass function is very difficult, even within the Local Group. Furthermore, at low redshift these small-scale modes are deep in the nonlinear regime and baryonic effects cannot be neglected, which mean that high resolution hydrodynamical simulations are necessary to make predictions and test observations. 

Original comparisons between observations of the dwarf satellite galaxies of the Milky Way to N-body simulations led to much attention being poured into the now well-known missing satellites \cite{missing_satellites} and too-big-to-fail problems \cite{TBTF1,TBTF2}. It now seems like these problems are more symptomatic of our assumptions about baryonic processes and galaxy formation on small scales being wrong or incomplete, rather than a shortcoming of the cold dark matter (CDM) scenario.
This motivates the use of a gravitational method for substructure detection. Within the Local Group, methods such as tidal streams \cite{Ngan:2013oga,2016ApJ...820...45C,Bovy:2015mda,2016MNRAS.463..102E} and looking at the motions of stars within the Milky Way \cite{Feldmann:2013hqa}, attempt to look for dark subhalos.
Beyond the Local Group, however, using images of strongly-lensed background galaxies or quasars is the only way, so far, of finding dark matter substructure. The main idea behind this method is that sufficiently massive substructures that lie close in projection to the lensed arcs or images can cause distortions that deviate from predictions based on a lens model with no substructure (also called a \textit{smooth lens} or \textit{macro model}). Additionally, the more numerous population of lower mass halos (expected in the CDM paradigm) can collectively cause perturbations to images that can be detected statistically. 

Several distinct methods have been proposed to quantify these distortions, both directly \cite{Mao,grav_imaging1,grav_imaging2,spat_res_spec1,spat_res_spec2} and statistically \cite{Hezaveh_powerspec,pcatlens,Birrer:2017rpp,Brewer:2015yya,dark_census,powerspec1,powerspec2,Brennan,mining_substructure}. Direct detection methods rely on having to model the smooth component of the lens galaxy before doing inference on the substructure abundance/properties. Lens modeling for real data is a complicated process that can take a long time for a single system, and different approaches to modeling a same system can yield different results (see e.g. \cite{2014MNRAS.442.2017V}). More importantly, mistakes in the macro model can further translate into false positive substructure detections, although steps can be taken to minimize the likelihood of this happening (see e.g. \cite{bells_2018}). Statistical methods such as the ones put forth in \cite{Hezaveh_powerspec,powerspec1,powerspec2,Brennan} also require an intermediate step of subtracting a smooth lens model from the data to look for correlations in the residuals, while others, such as \cite{pcatlens,Brewer:2015yya}, simultaneously infer the smooth model and subhalo population parameters. Most recently, Ref. \cite{mining_substructure} presented a very promising machine learning (ML) technique to explicitly evaluate the likelihood of an image given some subhalo population parameters (the fraction of dark matter mass contained in substructure and the subhalo mass function slope), marginalizing over all the smooth lens and individual subhalo parameters. This method is particularly promising because it can be used to combine any number of different images to put constraints on the subhalo population parameters. 

Just like Ref. \cite{mining_substructure} used machine learning to advance and speed up statistical searches for substructure, in this work our goal is to have a similar impact on direct detection efforts. One of our primary objectives is to gauge the feasibility of sidestepping the crucial smooth lens modeling step, both because of its potential to bias detections and because of its time cost. Furthermore, the process of inferring subhalo properties, even after a fit for the smooth lens model is obtained, is also computationally expensive and, more often than not, detailed analyses find no compelling evidence for the presence of substructure \cite{2014MNRAS.442.2017V,bells_2018}. In the coming years, however, hundreds or thousands of new strong lensing systems are expected to be found with optical imaging data \cite{LSST_lenses,2014MNRAS.439.3392P,Collett_2015} by experiments such as the Wide Field Infrared Survey Telescope (WFIRST), the Hubble Space Telescope (HST), the Large Scale Synoptic Survey (LSST), the Dark Energy Surves (DES), and \textit{Euclid}, vastly increasing the number of images that can be used for dark matter science\footnote{There are many other surveys, such as the interferometer Atacama Large Millimeter/submillimeter Array (ALMA), that are expected to find many new lenses as well, but in this work we focus more on optical imaging-type data instead of, for example, working with \textit{uv} visibilities.}. In an ideal world, each of these images could be analyzed individually but, in practice, having a fast method to find interesting candidates to focus resources on could accelerate the capacity of gravitational lensing-based methods to truly constrain dark matter properties. 

To this end, we present results using a convolutional neural network (CNN) \cite{cnn} to analyze strong lens images with varying sources, macro model parameters, and substructure populations to determine whether they are likely to contain detectable massive substructures in the vicinity of the Einstein radius of the lens. In particular, we do the inference directly on the (simulated) data, instead of first fitting a smooth model to the images and inferring the presence of dark substructures from the residual between the smooth model image and the data. More specifically, we seek to answer the question: \textit{what is the minimum mass that subhalos in the vicinity of the Einstein ring have to have for the neural network to identify a feature on the image as being due to ther presence instead of noise or the smooth model}? 
Note that it is not obvious that a neural network would be able to do this at all. Traditionally, classification done with NNs is based on some of the largest or most obvious features in images they are trained on, for example with the canonical MNIST \cite{mnist} or CIFAR10 \cite{cifar10} datasets. Indeed, previous works that have applied CNNs to do regression or classification directly on images in the context of strong lensing have mostly focused on macro model parameters. Refs. \cite{2018A&A...611A...2S} and \cite{2019MNRAS.487.5263D} proposed using them to identify images with strong lenses in photometric data; Refs. \cite{Hezaveh_cnn1} and \cite{PerreaultLevasseur_cnn2} applied this deep learning method to determine the parameters (and uncertainties) of the smooth lens model using optical imaging data; and \cite{Morningstar:2018ase} used CNNs and recurrent neural networks to extract smooth lens parameters from interferometric data. 

The focus of this work is fundamentally different: our goal is to address whether the image processing capability of a CNN is powerful enough to classify images based on minute differences (whether a perturber is present) even when the large-scale features of the images vary as well (the macro model parameters and sources vary from image to image). We therefore tackle the substructure detection problem as a binary classification task. Such an approach could be used to analyze any number of observed images as a filter to identify candidates that are likely to have a detectable subhalo somewhere in the image, such that traditional analyses can be carried out on them to determine substructure properties (i.e., masses and positions), while avoiding doing the same for images that would yield null detections. We emphasize that are not implying that images devoid of detectable substructure are not interesting, quite the opposite: null detections are a crucial ingredient to constrain the subhalo mass function and test the CDM paradigm (indeed, to date, constraints on subhalo properties using strong lens images are driven more by non-detections than detections). Rather, the point is that a pipeline such as the one we are suggesting in this work could determine what images have null detections in a fraction of a second instead of requiring detailed analyses for a much longer time. In essence, if the CNN's sensitivity to substructure were understood well enough, it could serve as a proxy for the detailed sensitivity function that has to be obtained on an image-by-image basis for gravitational imaging, e.g. \cite{detection_2010_mnras,vegetti_nature,bells_2018,2014MNRAS.442.2017V}, meaning that the the information from non-detections could be leveraged to constrain CDM as well. In this work, we focus in particular on galaxy-galaxy lensing systems, although this could be done for point-like sources such as quasars as well.

This paper is organized as follows. In Section \ref{sec:data_and_methods} we describe how we simulated strongly lensed images (Section \ref{subsec:data}), briefly review neural networks (Section \ref{subsec:cnn_review}), specify how we constructed our trainining/validation/test sets (Section \ref{subsec:tvt}), and present the specific architecture and optimization parameters employed in this paper (Section \ref{subsec:cnn}). In Section \ref{sec:results} we present our results and in Section \ref{sec:conclusion} we discuss the implications of this work and conclude.

\section{Data and Methods}\label{sec:data_and_methods}

\subsection{Simulating strongly-lensed images}\label{subsec:data}

We employ a neural network as a \textit{supervised} machine learning technique, meaning that, in order to learn, the algorithm requires the training data to be labelled. To train and evaluate the neural network we therefore use simulated strong lens images. We use the publicly available software package \texttt{lenstronomy} \cite{lenstronomy} to generate the images. We simulate images with $79 \times 79$ pixels that correspond to a field of view of $5.0'' \times 5.0''$, meaning they have a resolution of $0.06''$/pixel. 

\begin{figure*}
\includegraphics[width=\textwidth]{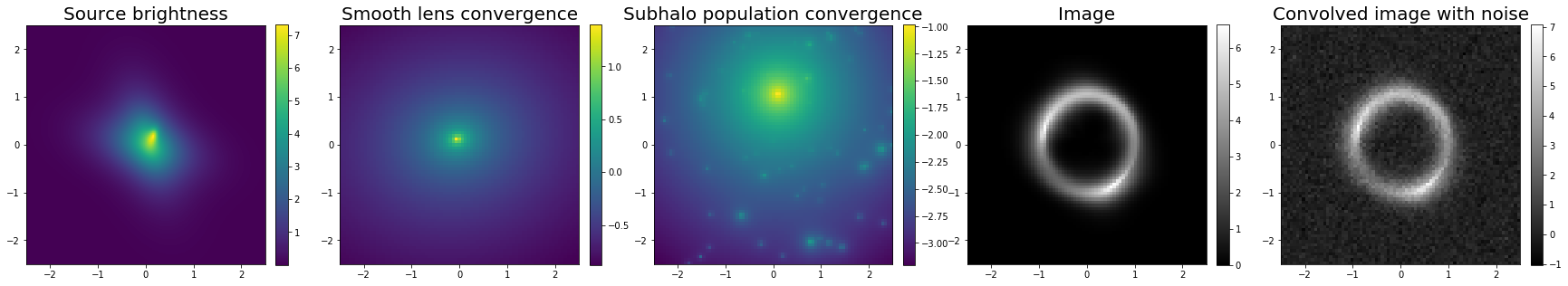}
\caption{\footnotesize{An example of the simulation pipeline (minus the negative mass sheet) for an image with a complete Einstein ring: the source brightness (first column), the convergence field of the smooth model in $\log$ units (second column), the convergence field of the subhalo population in $\log$ units (third column), the image resulting from the smooth model plus subhalos (fourth column), the image after convolving it with a PSF and adding noise (fifth column). The units of the panels are in arcseconds (''), each has a field of view of 5'' $\times$ 5'' and 79 $\times$ 79 pixels. The convergence field is simply the surface mass density normalized by the critical density for lensing, $\Sigma_{\rm crit} = c^2 D_{\rm os} / (4\pi G D_{\rm ol}D_{\rm ls})$, where $c$ is the speed of light, $D_{\rm os}$ the angular diameter distance from the observer to the source, $D_{\rm ls}$ between the lens and the source, and $D_{\rm ol}$ between the observer and the lens. The image brightness is in arbitrary units of surface brightness integrated over units of an angle squared. The source has $N_{\rm clumps} = 3$, the center of the lens is at $(x,y) = (-0.05",0.12")$, and its ellipticity is $(\epsilon_x,\epsilon_y) = (0.09,0.04)$, there are 52 subhalos and the highest mass is $m_{\rm high} = 9.9 \times 10^{9}$ M$_{\odot}$. The Gaussian PSF kernel has a size of 0.07'' and the images have Poisson noise corresponding to an exposure of 1000 seconds and 10\% white noise.} }\label{fig:sim_pipeline}
\end{figure*}

\begin{figure*}
\includegraphics[width=\textwidth]{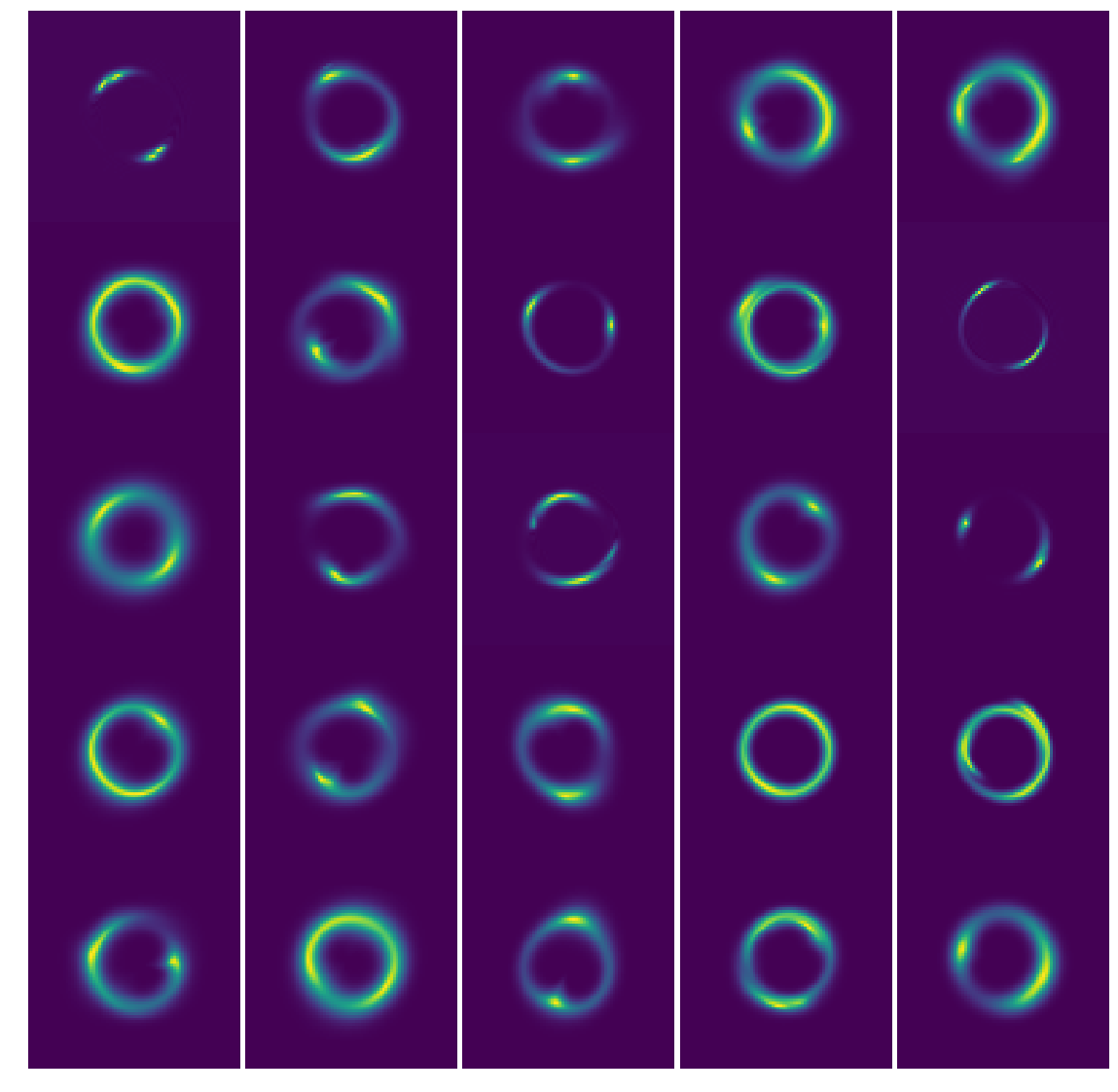}
\caption{\footnotesize{Examples of images generated with the simulation pipeline detailed in Section \ref{sec:data_and_methods}. The axes have been omitted for clarity, but all of these images correspond to a field of view of 5'' $\times$ 5'' (79 $\times$ 79 pixels).}}\label{fig:images}
\end{figure*}

Each image has five different ingredients (or three, in the case of macro-only images that contain no substructure): a smooth component, a stochastic population of subhalos, a negative mass sheet to compensate for the surface mass density added in subhalos, a simulated source of light, and instrumental effects and noise. Figure \ref{fig:sim_pipeline} shows an example of all the components that go into simulating an image that contains substructure (fourth column), except the mass sheet since it is just constant across the image: the source (first column), smooth model (second column), and subhalo population (third column). The fifth column shows what the image looks like once it is convolved with a point spread function (PSF) kernel and noise is added to it. More details about each of these steps are provided below and in the figure caption. Figure \ref{fig:images} shows several more examples of simulated images to illustrate that the width, completeness, and shape  of the Einstein ring, vary from image to image.

\subsubsection{Smooth lens model}

We model the smooth component of the main lens as a singular isothermal ellipsoid (SIE), with surface mass density given by \cite{SIE}:
\begin{equation}\label{eq:Sigma}
    \Sigma(\boldsymbol{r}) = \frac{ f \sigma_v^2}{2 G \sqrt{x^2 + f^2 y^2}},
\end{equation}
where $\boldsymbol{r} = \left( x,y \right)$ is the projected two-dimensional position on the lens plane, $\sigma_v$ is the velocity dispersion of the host along the line-of-sight, $f$ the axis ratio and $G$ the gravitational constant.

The SIE profile in \texttt{lenstronomy} is parametrized in an equivalent but slightly differently way: by the Einstein radius $\theta_{\rm E}$, which relates to the velocity dispersion by 
\begin{equation}
\sigma_v = \sqrt{\frac{\theta_{\rm E}}{4\pi}\frac{D_{\rm os}}{D_{\rm ls}}},
\end{equation}
where $D_{\rm os}$ is the angular diameter distance from the observer to the source, and $D_{\rm ls}$ the angular diameter distance from the lens to the source; and the $x$ and $y$ components of the ellipticity \{$\epsilon_x,\epsilon_y$\}, which are related to the axis ratio as
\begin{equation}
    f = 1-\sqrt{\epsilon_x^2 + \epsilon_y^2}.
\end{equation}
Finally, one can also specify the center of the lens \{$x,y$\}. 

In this work, we vary $x, y, \epsilon_x$ and $\epsilon_y$ from image to image and keep the Einstein radius fixed to $1.0''$\footnote{In reality, due to the dependence of the surface mass density on the axis ratio, see Eq. \eqref{eq:Sigma}, and the fact that the host has non-zero ellipticity, the Einstein radius of the images is never actually equal to one. Furthermore, the addition of substructure can change its value. See Section \ref{subsec:tvt} for a discussion of the steps we took to ensure that the Einstein radii of the images with substructure was consistent with that of the macro-only images.} for all the images (although its position does change due to the offset between the source and the center of the lens). This corresponds roughly to the typical angular size of galaxy-galaxy lenses. These parameters are chosen randomly from uniform distributions $U$ with ranges:
\begin{align}\label{eq:macro_params}
    x,y \sim U[-0.25'',0.25''] \\
    \epsilon_x,\epsilon_y \sim U[0,0.1]. 
\end{align}

The lenses are placed at $z_{\rm l} = 0.2$. Although it is not an intrinsic parameter of the host lens, the external shear (due to, for example, large-scale structure) is generally bunched into the parameters of the macro model. Here we do not include an external shear component, however, leaving this for future work.

\subsubsection{Subhalo population}

Our goal is to gauge the sensitivity of a neural network to perturbations caused by substructure. In the simplest case, we can therefore add a single subhalo to each image, changing its mass and position from image to image. Due to the abundance of substructure in the CDM paradigm, however, we expect there to be many subhalos in a projected area like the one under consideration in this paper. So we can instead add a stochastic population of subhalos, where the numbers, positions and masses of the subhalos vary from image to image. 

To populate our images we consider expectations from CDM and constraints from observations on the number of subhalos $N_{\rm sub}$ and the mass fraction in substructure, defined as $f_{\rm sub} = \sum_{i=1}^{N_{\rm sub}}m_i/M_{\rm host}$. Since these are functions of the range of subhalo masses considered, the host redshift, the host mass, etc., we compiled constraints for systems that are similar to our ensemble of lenses. From the high-resolution N-body simulation of a Milky Way-like halo ETHOS \cite{ETHOS}, Ref. \cite{powerspec2} finds that between redshifts of 0 and 0.5, there are be between 25 - 35 subhalos in a projected area corresponding to the field of view of the images we are considering, by averaging over many different lines-of-sight. Note, however, that this halo is about an order of magnitude smaller than the typical masses of massive elliptical galaxies, so this can be seen as a lower bound: Ref. \cite{Madau_2008} showed that a dark matter halo eight times more massive than a Milky Way halo can contain almost a factor of 2 more substructures with larger circular velocities. Ref. \cite{Xu:2013kna} finds that for a $10^{13}$ M$_{\odot}$ host, $0.1\% \lesssim f_{\rm sub} \lesssim 1\%$ between $10^{6} - 10^{11}$ M$_{\odot}$.
Using the lower limit from N-body simulations of $f_{\rm sub} = 0.3\%$ \cite{vialactea}, Ref. \cite{2009MNRAS.400.1583V} estimates that in the CDM paradigm, we can expect there to be $6.46 \pm 0.95$ substructures with masses between $4 \times 10^6 - 4 \times 10^9$ M$_{\odot}$ within an annulus of 0.6" centered on the Einstein radius \cite{Diemand:2006ik,Diemand:2007qr,Diemand:2008in}. 

Attempts to measure $f_{\rm sub}$ from substructure detections and non-detections have yielded values that vary considerably, but all seem to roughly agree with the expectations from N-body simulations. We cite in particular constraints where the lenses were at redshifts similar to the one we are using in this work. Ref. \cite{detection_2010_mnras} put a constraint of $f_{\rm sub} = 2.15^{+2.05}_{-1.25}\%$ or $f_{\rm sub} = 2.56^{+3.26}_{-1.50}\%$ between $4 \times 10^6 - 4 \times 10^9$ M$_{\odot}$, depending on the choice of prior, with a single lens at $z_{\rm l} = 0.222$. 
More recently, using 11 SLACS lenses, Ref. \cite{2014MNRAS.442.2017V} put a constraint of $f_{\rm sub} = 0.76^{+2.08}_{-0.52}\%$ or $f_{\rm sub} = 0.64^{+0.8}_{-0.42}\%$, depending on the choice of prior, between $4 \times 10^6 - 4 \times 10^9$ M$_{\odot}$ in a sample of lenses with mean $\langle z_{\rm l}\rangle = 0.2$. Using the same sample of lenses, Ref. \cite{Vegetti:2018dly} finds a higher value of $f_{\rm sub} < 8.7\%$ (68\% C.L.) between $10^5 - 10^{11}$ M$_{\odot}$, which the authors attribute to a different definition of substructure mass and mass limits, and a different shape of the substructure mass function.

To populate the host lenses with substructure we therefore consider the following points. First, we certainly want to test the networks on images that are considered ``realistic" by the above guidelines. However, as can be glimpsed by the plethora of different values we gathered above, it is not necessarily clear what ``realistic" means. Furthermore, the small number of actual detections and systems that have been analyzed to date means that we do not yet have a sufficient grasp of what these subhalo populations actually look like in real lenses, outside the idealized scenarios of N-body simulations that do not take into account the impact of baryonic physics. Therefore, we also want to gauge the performance of the network on a broader range of types of subhalo populations, to see how the network could fare if real subhalo populations deviate either slightly, or significantly, from the expectations based on N-body simulations.

Taking all these factors into consideration, we devise two different schemes for populating lenses with substructure. In our first approach, which we call the $N_{\rm sub}$-bound approach, we impose a constraint on the number of subhalos that lie in the area covered by the image, and vary the highest subhalo mass $m_{\rm high}$ (Section \ref{sec:contrain_nsub}). This has the consequence of having images with very different values of $f_{\rm sub}$, since it will be highly dependent on the value of $m_{\rm high}$ in an image. In the second approach, which we call $f_{\rm sub}$-bound, we instead constrain the value of $f_{\rm sub}$. This has the consequence of having very different numbers of subhalos depending on what $m_{\rm high}$ is. For instance, for $m_{\rm high} << f_{\rm sub}M_{\rm host}$, thousands of subhalos are required to satisfy the bound on $f_{\rm sub}$, while for $m_{\rm high} \lesssim f_{\rm sub}M_{\rm host}$, a handful of subahlos saturate it. 

These two approaches, together with the single-subhalo case, are complimentary and shed light on different aspects of the network's sensitivity. For example, we can understand whether it is more sensitive to a larger number of lower mass subhalos, or a lesser number of more massive halos. For a given value of $m_{\rm high}$ we can also understand how the presence (or absence) of other lower mass perturbers affects the network's sensitivity. Furthermore, it allows us to explore the network's behavior among a broad range of different subhalo population characteristics. 

The perturbers are placed within the lens itself, as opposed to considering line-of-sight halos that lie in the vicinity of the lensed arcs in projection. All the subhalos are always modeled with Navarro-Frenk-White (NFW) \cite{nfw} profiles, and their concentrations fixed to $c=15$. This choice is intended to be representative of the average concentration one might expect for substructure. In reality, concentration is a non-trivial function of mass and redshift. Concentration-mass relations extracted from cosmological N-body simulations show that it is a decreasing function of mass and an increasing function of redshift. In the mass ranges we are considering, the concentration is predicted to be $\gtrsim 10$. When considering subhalos, however, there is the added complication that they are tidally stripped as they move within their host's potential, so the concentration is not necessarily always well defined. Instead, subhalo profiles are sometimes characterized with an analogous parameter which is the ratio between the tidal radius and the scale radius. Using the same phenomenological equations as Ref. \cite{powerspec1}, we find that this parameter is roughly $\lesssim 29$ (with the upper bound corresponding to subhalos at the outskirts of the host), and an average of $\approx 20$. Because the tidal radius is smaller than the virial radius \cite{Diemand:2007qr} the corresponding values of the concentration are expected to be somewhat smaller than this.

\bigskip 

\paragraph{Single subhalo}\label{sec:one_subhalo}

We first draw a mass from a log-uniform distribution between $10^8 - 10^{11}$ M$_{\odot}$. \texttt{Lenstronomy} parametrizes the deflection due to NFW profiles using the scale radius in angular units and the radial deflection angle at the scale radius, and to convert physical NFW masses to these parameters we need to specify a cosmology and source and lens redshifts. We place the source at $z_{\rm s} = 0.6$, the lens at $z_{\rm l}=0.2$, and use the \textit{Planck} 2015 cosmology \cite{Planck2015}. 

We allow the subhalo to be at any position in the image that satisfies two constraints:
\begin{enumerate}
\item It has to lie within $0.25''$ of the Einstein radius of the host, i.e. $|\theta_{\rm E} - r_{\rm high}| \leq 0.25''$, where $r_{\rm high} = \sqrt{x_{\rm high}^2+y_{\rm high}^2}$. 
\item The intensity at $r_{\rm high}$ must be greater than or equal to a minimum intensity threshold, unique to each image, determined by generating a macro-only image with the same macro model parameters, masking the annulus encompassed by $\theta_E \pm 0.35''$, and obtaining the maximum of the masked image.
\end{enumerate}
The reason behind these constraints is that direct detection methods only have sensitivity to substructure in the vicinity of the Einstein ring or arcs in the image (e.g. \cite{grav_imaging1,grav_imaging2}). Therefore, if we want to know whether perturbations caused by subhalos of a given mass can be detected by the CNN, the substructure with that mass must be close to the ring/arcs. In particular, the reason for the second constraint is to make sure that, in images where the ring is largely incomplete, the subhalo still lies near an area with non-negligible intensity. 

\bigskip 

\paragraph{Constraining the number of subhalos}\label{sec:contrain_nsub}

In the $N_{\rm sub}$-bound approach we generate images where the number of subhalos is drawn from a Normal distribution with mean $\mu = 60$ and standard deviation $\sigma = 15$. 
Once $N_{\rm sub}$ is drawn, we sample $N_{\rm sub}$ masses from a subhalo mass function consistent with CDM, taken to be a power law with slope $\beta=-1.9$ \cite{Aquarius} between $m_{\rm min} = 10^6$ M$_{\odot}$ and $m_{\rm high}$, with $m_{\rm high}$ anywhere between $10^8 - 10^{11}$ M$_{\odot}$. We again convert the masses into the the scale radius in angular units and the radial deflection angle at the scale radius as required by \texttt{lenstronomy}, using the same cosmology and source and lens redshfits as above.
 
 All the subhalo positions are chosen randomly to lie within the full area of the image\footnote{Due to projection effects, and the fact that the area probed by lensing transverse to the line-of-sight is very small, the subhalo distribution is essentially isotropic in the region of interest. See e.g. \cite{powerspec2}.}, except for that of the most massive subhalo. The position of the most massive subhalo $r_{\rm high}$ is modified to obey the same two constraints as those detailed in Section \ref{sec:one_subhalo} above. 

\bigskip 

\paragraph{Constraining the fraction of mass in substructure} \label{sec:contrain_fsub}

In the $f_{\rm sub}$-bound approach, we fix the mass fraction in substructure to be $1 \pm 0.05 \%$. For a given value of the highest subhalo mass $m_{\rm high}$, we generate draws of the subhalo mass function until the $f_{\rm sub}$ constraint is satisfied. In the cases where $m_{\rm high}$ is high enough that it already saturates the bound on $f_{\rm sub}$, we instead draw a different set of masses from the subhalo mass function that obeys the bound, and append the most massive subhalo. Alternatively we could have added a single subhalo to the image in this regime, but since we already had a dataset comprised of images with a single subhalo, we opted for this alternative approach here. In this way, we could see if there is any difference in the network's sensitivity in this regime due to an additional population of low mass subhalos. 

The position of the most massive subhalo in a given image is again constrained by the two conditions described in Section \ref{sec:one_subhalo}.

\bigskip 

More details on some of the relevant properties of the subhalo populations in the images used for training/validating/testing are provided in Appendix A.

\subsubsection{Negative mass sheet}

Since we fix the Einstein radius of the host to 1", we add a negative mass sheet to the substructure images to ensure that the convergence in the image is the same as for the macro-only images (since the Einstein radius depends on the convergence). The convergence field is simply the surface mass density normalized by the critical density for lensing, $\Sigma_{\rm crit} = c^2 D_{\rm os} / (4\pi G D_{\rm ol}D_{\rm ls})$, where $c$ is the speed of light, $D_{\rm os}$ the angular diameter distance from the observer to the source, $D_{\rm ls}$ between the lens and the source, and $D_{\rm ol}$ between the observer and the lens. For each image with substructure we generate a macro-only image with the same macro-model parameters, find the total difference in convergence between the two images and add a negative convergence field that cancels this difference. All the results we present correspond to substructure and macro-model images that have the same effective Einstein radius, defined as the radius from the host center at which the convergence decreases below one.

\subsubsection{Source}

An image's sensitivity to substructure, measured as the surface brightness change $\delta I_{\rm sub}$ due to a potential perturbation caused by a subhalo $\delta\psi_{\rm sub}$, is proportional to the gradient of the source $\nabla S$ \cite{grav_imaging1,grav_imaging2,Cyr-Racine_2018}:
\begin{equation}
    \delta I_{\rm sub}(\boldsymbol{y}) = - \nabla S(\boldsymbol{x})|_{\boldsymbol{x} = \boldsymbol{y} - \nabla\psi_0(\boldsymbol{y})} \cdot \nabla \delta\psi_{\rm sub}(\boldsymbol{y}),
\end{equation}
where $\boldsymbol{y}$ are the coordinates on the image plane and $\boldsymbol{x}$ the coordinates on the source plane. The gradient of the source brightness distribution evaluated on the source plane is translated into the image plane with the lens equation evaluated with the smooth component of the lens $\psi_0$. This is why highly structured sources, for example dusty star-forming galaxies that are very clumpy, are considered prime candidates to find subhalos (see e.g. \cite{spat_res_spec2}). 

Here we focus on extended sources instead of point-like sources like quasars. We simulate sources with some degree of structure but not so much that it would be unlikely to be resolved by typical optical imaging surveys (i.e. much less structured than the simulated sources in Ref. \cite{spat_res_spec2} used to forecast the sensitivity of ALMA). We model the source $S$ as one or more discrete but very close-by clumps of light $s_i$ each modelled as a S\'ersic ellipse, determined by five parameters for the $i$th clump: the amplitude of the intensity $I_{i}$, the half-light radius $R_{{\rm ser},i}$, the Sersic index $n_{i}$, and the $x$ and $y$ components of the ellipticity \{$\epsilon_{x,i}$,$\epsilon_{y,i}$\}. 

The source is different in each image. For a given image, the number of clumps is drawn from a uniform distribution $N_{\rm clumps} \sim U[1,4]$. For each clump, the amplitude of the intensity is always fixed to unity (in arbitrary units of surface brightness integrated over units of an angle squared), and the remaining four parameters of a Sersic ellipse are drawn from the following uniform distributions:
\begin{align}
R_{{\rm ser},i} \sim  U[0.1 \text{kpc},1 \text{kpc}] \\
\epsilon_{x,i},\epsilon_{y,i} \sim  U[-0.5,0.5]. \\
\end{align}
$\epsilon_{x,i}$ and $\epsilon_{y,i}$ are subject to the additional constraint that the magnitude of the ellipticity is $\epsilon_{i} = \sqrt{\epsilon_{x,i}^2 + \epsilon_{y,i}^2} \leq 0.4$. If there is a single clump, it is chosen to lie at the center of the image. For $N_{\rm clumps} > 1$, the relative positions of the clumps are drawn from a multivariate Normal distribution with mean $\boldsymbol{\mu} = (0,0)$ and covariance matrix with diagonal entries $\sigma_{xx}^2 = \sigma_{yy}^2 = 0.01$ and off-diagonal entries $\sigma_{xy}^2 = \sigma_{yx}^2 \sim U[-0.25,0.25]$. The final source $S$ is a sum of all the individual clumps, $S = \sum_{i=0}^{N_{\rm clumps}} s_i$. As mentioned above, we place the source at $z_{\rm s} = 0.6$. 

\subsubsection{Instrumental effects and noise}

After the lensed image has been generated, it is convolved with a Gaussian point spread function (PSF) kernel with a full-width-half-max (FWHM) of $0.07''$ (roughly equivalent to that of HST). Then, Poisson shot noise for an exposure of 1000 seconds and Gaussian noise with a standard deviation given by some fraction $p$ of the mean signal in the Einstein ring/arcs are added to the image; we showcase $p = \{0.01,0.1,0.33\}$. 

\begin{figure*}
\includegraphics[width=1.0\textwidth]{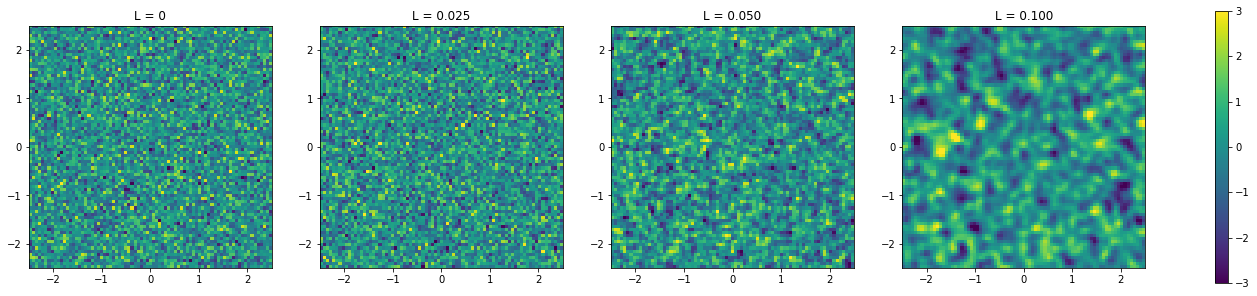}
\caption{\footnotesize{Examples of noise with varying correlation length $L$, obtained using the covariance matrix in Eq. \eqref{eq:covar}. The standard deviation was fixed to $\sigma=1$ for this Figure, and all the panels share the same colorbar (units are arbitrary). Starting from the left, the first panel corresponds to no correlation ($L=0$). The second panel corresponds to a very small correlation length ($L=0.025$), which is indistinguishable from the uncorrelated case. The third and fourth columns correspond to $L=0.05$ and $L=0.1$, respectively. The increase in correlation between the pixel values as $L$ increases is readily apparent. }}\label{fig:noise}
\end{figure*}

Generally, the Gaussian noise added to simulated lensed images in the literature is uncorrelated and independent in each pixel. However, real data could have more complicated, correlated noise among nearby pixels due to, for example, drizzling \cite{drizzling}. We thus also test the performance of the NN when the noise added to the image is correlated. We expect that this will degrade the classification accuracy, since correlated noise could replicate the effect of subhalos more closely. We use a Gaussian Process with a squared exponential kernel $K$,
\begin{equation}\label{eq:covar}
    K(x,x') = \sigma^2 \exp\left(-\frac{(x-x')^2}{2L^2}\right),
\end{equation}
to generate the correlated noise. $\sigma$ is the standard deviation and $L$ is the lengthscale, which determines the distance over which the pixels are correlated. We use the same value of $\sigma$ as in the uncorrelated case ($L=0$) and vary $L = \{0.05, 0.1\}$. To speed up sampling from this multivariate Gaussian distribution we use the reparametrization trick \cite{reparametrization} with the Cholesky decomposition of the covariance matrix. Figure \ref{fig:noise} shows examples of the noise with varying values of $L$. 

\subsection{Neural Networks}\label{subsec:cnn_review}

For readers unfamiliar with neural networks, we provide here a brief, high-level overview of how they work.

Neural networks are an extremely powerful tool when one has a dataset that consists of input-output pairs, $(\boldsymbol{X}_{\{i\}},\boldsymbol{y}_{\{i\}})$ and wants to be able to obtain outputs $\boldsymbol{\hat{y}}_j$ given inputs $\boldsymbol{X}_j$ for $j \notin \{i\}$. The outputs are also referred to as (class) labels in the case of classification. Neural networks simply act as extremely complicated functions $f_{\rm NN}$ that are taught how to map an input $\boldsymbol{X}_k$ to an output $\hat{\boldsymbol{y}}_k$, $\hat{\boldsymbol{y}}_k = f_{\rm NN}(\boldsymbol{X}_k)$. 

Neural networks are arranged into \textit{layers}, and each layer contains a number of \textit{neurons}. There are two different types of layers that are relevant to this work. The first type are \textit{fully-connected}, also called dense, layers. These are one-dimensional layers where the neurons in a given layer are connected to all the neurons in the previous and following layers. The other type of layers are \textit{convolutional} layers, which tend to be two- or three-dimensional. In these layers, the inputs are convolved with a \textit{filter} that is slid over the neurons. Regardless of the type of layer, each neuron in a layer takes a linear combination of its inputs and applies a nonlinear function to them (called \textit{activation function}). 

The process of learning involves feeding the neural network many thousands (or more) of samples $\boldsymbol{X}_i$ for which their true label $\boldsymbol{y}_i$ is known, and optimizing all the parameters in the network (e.g. the weights and biases used to make linear combinations of inputs at each neuron) to minimize the \textit{loss function}, which quantifies the difference between the predicted $\hat{\boldsymbol{y}}_i$ by the network and the true value $\boldsymbol{y_i}$.

 In general, the optimization is done numerically with \textit{stochastic gradient descent} (SGD), or some variant of it. The general idea is to use the chain rule to find the gradient of the loss with respect to every parameter in the network (called \textit{backpropagation} in the ML jargon), and update the parameters after each iteration to minimize the loss until convergence is reached. 

\subsection{Training, validation and test sets}\label{subsec:tvt}

 To convert simulated images into training/validation/test sets that can be fed into a neural network, they need to be given class labels $C$. Since we are trying to understand down to what subhalo mass is the CNN sensitive, we turn the problem into a binary classification task. We train a neural network on images with no substructure (macro-only images), labeled with a zero ($C=0$), and images with substructure, with the highest subhalo mass anywhere between $m_{\rm high} = 10^8 - 10^{11}$ M$_{\odot}$, all labeled with a one ($C=1$). 

We carry out this test with datasets that have different subhalo populations, different levels of noise, and different amounts of noise correlation. The number of samples in the training/test/validation sets can vary slightly from table to table and row to row, but we ensure that there are at least $10^5$ images for each. For cases in which there is an uneven number of images for a given class $N_{{\rm samples},C=i}$, we compensate by weighing the loss function by the inverse of the fraction of training samples in a given class $f_{C=i}$:
\begin{equation}
    w_{C=i} = \frac{1}{f_{C=i}} = \frac{N_{\rm samples}}{N_{{\rm samples},C=i}},
\end{equation}
for $i=\{0,1\}$. When a given class has images with substructure, we ensure that there are an equal number of images with $10^8 < m_{\rm high}/M_{\odot} < 10^9$ and $10^9 < m_{\rm high}/M_{\odot} < 10^{11}$. 
We use 80\% of the images for training, and 10\% each for validating and testing. Furthermore, the training set is always augmented on-the-fly: each image is rotated by a random angle before going through the network, meaning that the network never sees exactly the same image twice. This helps prevent overfitting and also teaches the CNN rotational invariance. 

\subsection{CNN architecture and optimization strategy}\label{subsec:cnn}

We used \texttt{pytorch} \cite{pytorch} to implement the CNN. The results presented in this work are the result of a non-exhaustive grid search (see below for details) carried out using the $N_{\rm sub}$-bound training/validation sets that had 1\% uncorrelated noise.
Our goal is to have our results serve as a proof-of-concept, showing that CNNs can become a valuable tool to help tackle an extremely complicated problem, not to spend many extra GPU hours squeezing every last point of accuracy,
especially since our simulated images are not geared to replicate any one particular experiment. In reality, if one wanted to apply a pipeline like the one we are suggesting here to images taken by one (or several different) experiment(s), then a more exhaustive grid search could be carried out to improve the accuracy further.

Along these lines, we emphasize that once a good CNN architecture was found using this training/validation set, the same architecture was used for the other training sets considered. For example, the training set we used to do the grid search had images with 1\% noise, but when we train a network on images with a different amount of noise, we do not carry out a new grid search to re-optimize the network architecture. It is likely that the architecture could be fine tuned further to improve the results for the different training/validation sets. 

The grid search consisted of running the networks for 100 epochs and using the accuracy over the validation set to rank the networks' performance.
The parameters that were fixed or varied in the grid search are as follows. The network was forced to have two convolutional layers, each with varying filter size, stride, and number of channels. There was no zero padding, batch normalization was imposed, but maxpooling was optional. The number and width of fully connected layers was allowed to vary, from zero to five layers, and 25 to 100 nodes each. The network weights were initialized using a Normal Xavier initialization \cite{xavier}, whereby the weights are drawn from a Normal distribution with mean zero and variance $\sigma^2$ determined by $\sigma = \sqrt{2/(N_{\rm in} + N_{\rm out})}$, where $N_{\rm in}$ ($N_{\rm out}$) is the number of input (output) neurons. Preliminary tests indicated that augmenting the data did an excellent job of preventing overfitting, so we did not implement dropout in any layer nor a regularization term in the loss function. The activation function for all layers (except for the last layer) was a ReLU function. The final outputs were passed through a softmax function, which transforms them into probabilities, by restricting them to lie between 0 and 1 and together sum to 1. We use these to assign classes: for a given image, if the probability of class 0 (1) is greater than the probability of class 1 (0), then it is assigned to class 0 (1). The loss function we used was the negative log likelihood applied to the outputs of the softmax function, and we trained the CNN using the Adam optimizer. We set the learning rate to 0.001 and did not use a learning rate scheduler. The batch size was fixed to 64 samples per graphics processing unit (GPU), and 4 NVIDIA Tesla 2xK80s GPUs were used to train the network. 

\bigskip 

\begin{minipage}{\linewidth}
  \footnotesize
 \centering
 \captionof{table}{Network architecture.} 
 \hspace{-0.5cm}\begin{tabular}{ | c | c | c | c | }
\hline
\textbf{Number} & \textbf{Layer Type} & \textbf{Features} & \textbf{Dimension} \\ \hline \hline
1 & 2D Convolution & Filter size: 7  & Input : $1 \times 79 \times 79$\\
& & Depth: 16 & Output : $16 \times 37 \times 37$ \\ 
& & Stride: 2 & \\ 
& & Maxpool: False & \\  \hline
1& 2D Convolution & Filter size: 4 & Input : $16 \times 37 \times 37$\\
& & Depth: 16 & Output : $16 \times 12 \times 12$ \\ 
& & Stride: 3 & \\ 
& & Maxpool: False & \\  \hline
1 & Fully-connected & & $ 2304$ \\ [0.2cm] \hline
4 & Fully-connected & & 75 \\ [0.2cm] \hline 
 1 & Fully-connected & & 2 \\ [0.2cm]
\hline
\end{tabular}\par \label{tab:architecture}
\scriptsize{This network has 195,103 trainable parameters. Note that we have omitted the batch size in the dimensions of the inputs/outputs for clarity.} 
\end{minipage}

\bigskip 

\begin{figure*}
    \centering
    \begin{subfigure}[b]{0.49\textwidth}
        \includegraphics[width=\textwidth]{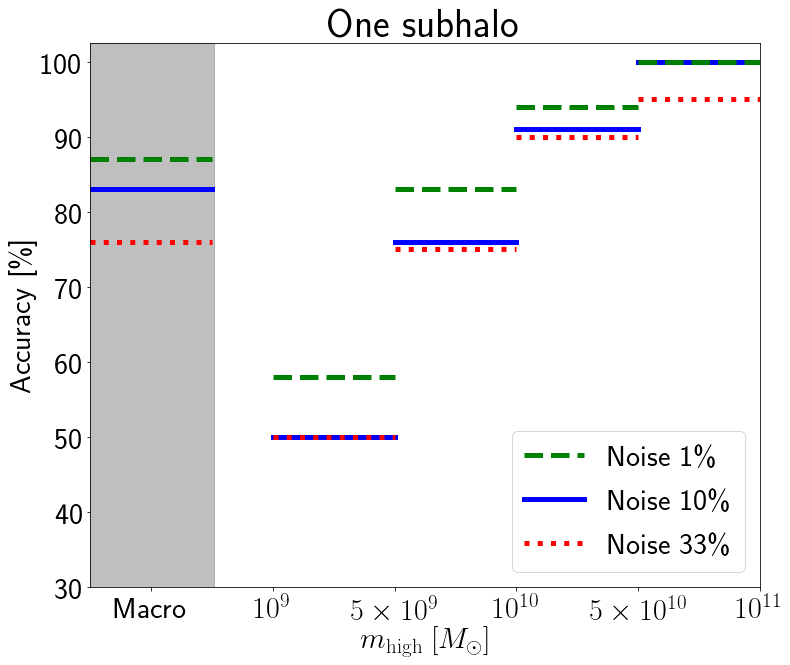}
    \end{subfigure}
    \begin{subfigure}[b]{0.49\textwidth}
        \includegraphics[width=\textwidth]{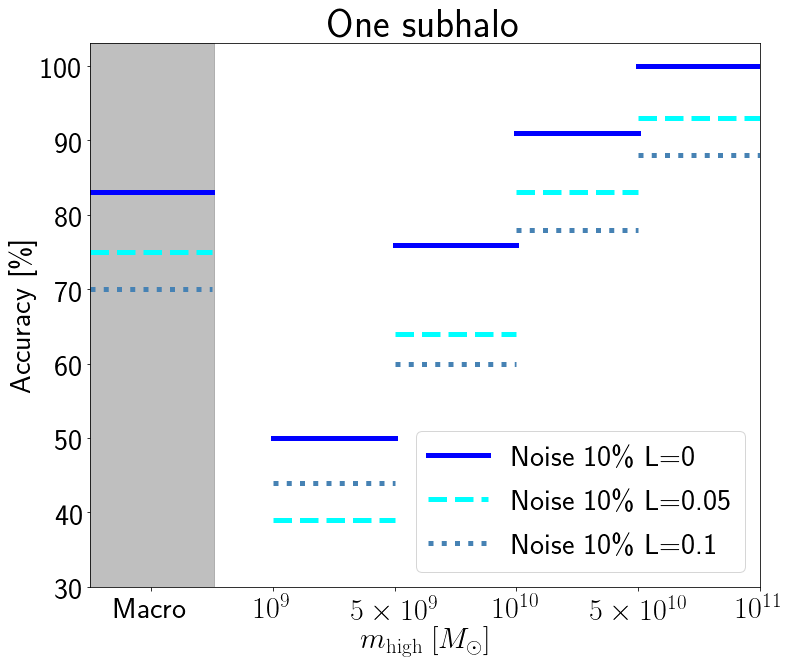}
    \end{subfigure}
    \begin{subfigure}[b]{0.5\textwidth}
        \includegraphics[width=\textwidth]{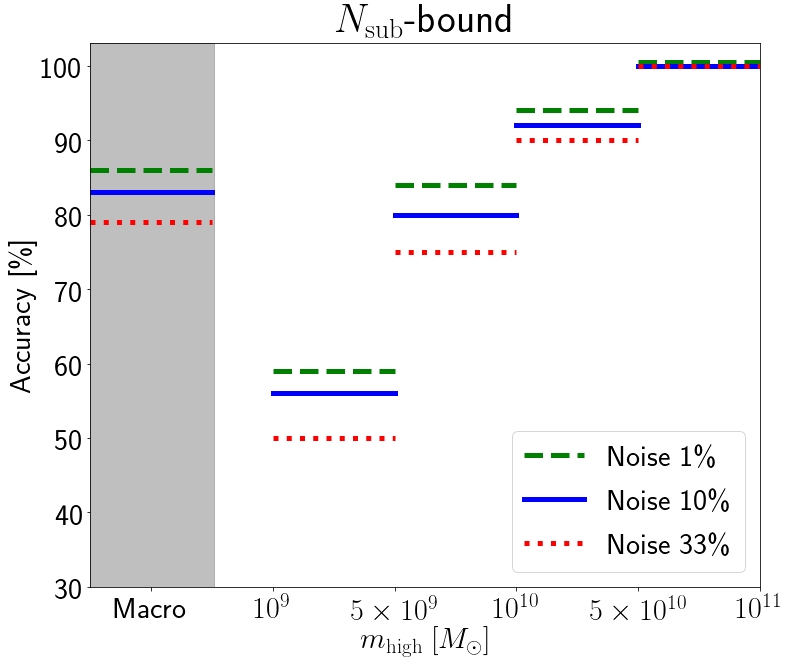}
    \end{subfigure}
    \begin{subfigure}[b]{0.5\textwidth}
        \includegraphics[width=\textwidth]{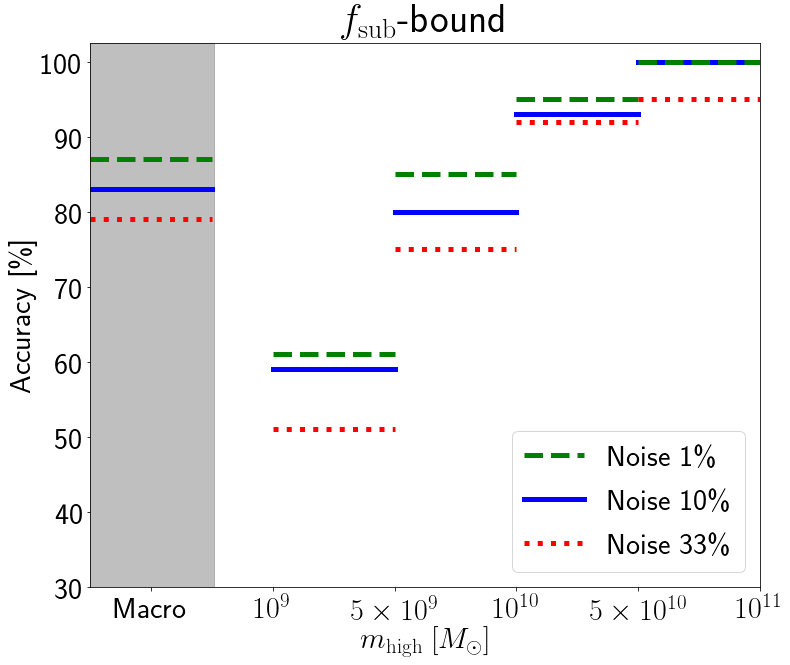}
    \end{subfigure}
    ~
\caption{\footnotesize{Percentage classification accuracy for the one-subhalo (top), $N_{\rm sub}$-bound (middle) and $f_{\rm sub}$-bound (bottom) test set images, as a function of the subhalo mass and for different levels of noise. In the middle panel, the accuracy line for the 1\% case in the mass bin $(0.5-1) \times 10^{11}$ $M_{\odot}$ has been shifted upwards slightly for it to be visible, because the classification accuracy for all three levels of noise was the same. The top right panel shows the classification accuracy when the level of noise is fixed to 10\% and instead the correlation lengthscale (as defined in Eq. \eqref{eq:covar}) is increased. All images have the same effective Einstein radius.}}\label{fig:accuracy_fiducial}
\end{figure*}

During the grid search we found that several different architectures had the same classification accuracy after 100 epochs, so out of these we picked the one with the least amount of parameters (the number of parameters varied by more than two orders of magnitude). Opting for the minimal number of parameters is beneficial to speed up training and prevent overfitting. Table \ref{tab:architecture} shows the network architecture that we ultimately used. The network has 195,103 parameters. 

\section{Results}\label{sec:results}

All the numbers given in this section correspond to having trained the network for 100 epochs with a learning rate of $10^{-3}$, and an additional maximum of 30 epochs with a learning rate of $10^{-4}$. All these results correspond to substructure and macro-only images that have the same effective Einstein radius.

Generally, the results for classification networks are given in terms of the overall accuracy. In this context, however, that number does not give us any insight into the NN's capacity to identify subhalos based on their mass. Furthermore, it is going to be strongly dependent on the distribution of $m_{\rm high}$ in the test set: we know that images with low $m_{\rm high}$ will be misclassified, since to the network they will be indistinguishable from macro-only images, while those with high $m_{\rm high}$ can be classified with more ease. This is true also of the ROC curve, which is oftentimes used to gauge the performance of a binary classifier. Due to this, we instead consider separately the accuracy for images with substructure and for macro-only images. In particular, for images with substructure we quantify the classification accuracy as a function of the highest subhalo mass.

The results for all three datasets are shown in Figure \ref{fig:accuracy_fiducial}. The mass range between $10^8 - 10^9$ M$_{\odot}$ is omitted because its accuracy lies below 50\%. Typically, in binary classification problems, the worst job a classifier can do is having around a 50\% accuracy, since this corresponds to classifying randomly. Here, we can see that for low values of $m_{\rm high}$, the accuracy in fact lies well below 50\%. This reflects the fact that, to the network, these images are indistinguishable from macro-only images: the perturbations created by subhalos with these masses are not significant enough to create features the network can identify and distinguish from features of the source and/or macro model (or the noise, in the case of images with noise). 

It is immediately apparent that the results are very similar in all three data sets for the three levels of uncorrelated noise considered.  This shows us that the network is not being aided significantly by the presence of additional lower mass perturbers in the $N_{\rm sub}$-bound and $f_{\rm sub}$-bound data sets, and the classification is still mainly driven by the single most massive subhalo. Furthermore, while in the lowest mass bin shown the accuracy consistently surpasses 50\% for low noise, we can see that subhalos have to be quite massive in order to have a classification accuracy considerably greater than 50\%: the mass bin $(0.5 - 1) \times 10^{10}$ M$_{\odot}$ is where we start seeing accuracy $\geq 75\%$. Unsurprisingly, increasing the level of noise decreases the classification accuracy.  

The addition of correlated noise, shown in the top right panel, decreases the accuracy in all the mass bins and for the macro-only images. For large correlations, the perturbers have to be somewhat more massive before they can be identified with $\geq 75\%$ accuracy, between $(1 - 5) \times 10^{10}$ M$_{\odot}$. 

As a null test, we attempted to train a network using the same images as those in the single-subhalo dataset but without having imposed the positional constraint on the most massive subhalo, meaning it was allowed to lie anywhere in the image. Since the area of the full image is much larger than the area covered by the lensed arcs/rings, it is much more likely for the highest mass subhalo to lie somewhere where it cannot have a significant impact. Our expectation was that the network should not be able to learn to distinguish these from the macro-only images. Indeed this is what we observed. 

\section{Discussion and Conclusion}\label{sec:conclusion}

In this work we have set out to explore the usability of convolutional neural networks $-$ a machine learning technique whose image recognition capability has achieved astonishing results in many different fields over the last few years $-$ to infer the presence of dark matter substructure in strong lens images directly, without having to carry out the inference at the level of the residual between a smooth model image and the observed image. Model fitting is troublesome in its own right: modeling choices and conventions for smooth lenses vary drastically in the literature, and it is possible to find quite different results for a same lens. Since finding a model for the smooth lens is (generally) a prerequisite to finding substructure, any mistakes in the model-fitting step can trickle down into inference on the presence of substructure and its properties. Methods such as gravitational imaging mitigate the likelihood of such an eventuality by carrying out a pixel-based reconstruction of the potential instead of relying solely on an analytic fit to minimize the residual between a smooth model and a model that also has one (or more) clump(s). Such analyses have the additional advantage of being Bayesian; however, they are very computationally expensive and time consuming. Most importantly, more often than not they result in null detections. 

With the expectation of thousands of new high-quality strong lens images becoming available in the near future, these factors thus motivate the development of fast, model-independent techniques to analyze strong lens images and find substructure, or come up with principled ways of choosing how to divert resources to where they can have the largest scientific impact. To this end, we train a CNN to classify images based on whether they have substructure or not. We emphasize that this classification problem is non-trivial because we are asking the network to classify images based on minute features while introducing huge variations in the large-scale characterstics from image to image, since the macro and source model parameters (which are highly degenerate with the substructure) vary from image to image. We believe that phrasing the substructure problem in this binary way could be advantageous because it means that the images that are found to contain no detectable substructure (i.e. classified as indistinguishable from macro-only images) will not have to see more resources diverted towards them just to return null detections. Instead, resources and time can be spent analyzing images that the CNN finds are likely to have detectable substructure. If the network's sensitivity is understood sufficiently well (analogously to the sensitivity function in gravitational imaging), detections could be leveraged with non-detections to constrain the cold dark matter scenario. 

We found, however, that subhalos have to be very massive, $m_{\rm sub} \gtrsim 5 \times 10^9$ M$_{\odot}$, in the vicinity of the Einstein ring in order to be recognized with an accuracy $>75\%$. Furthermore, the sensitivity does not seem to improve noticeably due to the presence of a larger population of lower mass subhalos, meaning that the classification is essentially driven by single very massive perturbers.

Comparisons between the sensitivity of this network and that of different methods to detect substructure is not straightforward, since sensitivity to substructure is a function of many different variables, such as the image resolution, the noise, and the source structure. Furthermore, we have not taken into account additional complicating factors in our simulated data, such as the host (or other sources of) brightness, or cosmic rays/bad pixels. 

However, we can attempt to put the capacity of this CNN into context, keeping in mind the simplified nature of our simulated data. In Refs. \cite{grav_imaging1,grav_imaging2}, the gravitational imaging technique applied to HST-like simulated images (with a resolution of 0.05''/pixel and a signal-to-noise of \textit{at least} 3 per pixel) was shown to have a sensitivity to subhalo masses as low as a few times $10^8$ M$_{\odot}$ for an NFW profile when the substructure is on the Einstein ring, and quickly increases with distance from the lensed images.  

In terms of actual detections, to date two systems with compelling evidence for substructure have been found with gravitational imaging. One of these, SDSSJ0946+1006, was found to have a subhalo with mass $(3.51 \pm 0.15) \times 10^9$ M$_{\odot}$ \cite{detection_2010_mnras}, while the other, JVAS B1938+666, was found to have a subhalo with mass $(1.9 \pm 0.1) \times 10^8$ M$_{\odot}$ \cite{vegetti_nature}. In both cases the subhalos were modelled as truncated pseudo-Jaffe profiles \cite{Dalal:2001fq} to obtain mass estimates; mass estimates done with NFW profiles tend to recover masses that can be significantly higher (for instance around $\sim 10^{10}$ M$_{\odot}$ for the $3.51 \times 10^9$ M$_{\odot}$ Pseudo-Jaffe subhalo in SDSSJ0946+1006 \cite{Vegetti:2018dly}). 

It therefore seems like the sensitivity of this CNN might be sufficient to find the perturber in SDSSJ0946+1006; it is less likely that it could find the one in JVAS B1938+666. It is worth keeping in mind, however, that the values for the accuracy as a function of subhalo mas cited in this work are the \textit{true} subhalo masses, while direct detection efforts are sensitive only to the \textit{effective} subhalo mass: Ref. \cite{effective_mass} showed that the true subhalo mass can be biased by up to an order of magnitude higher than what is actually measured with strong lensing, the effective mass, meaning it is possible that claimed substructure detections actually have higher true masses than the numbers that are given. 

If one did want to apply a method such as this to real data, much work would have to be done to understand the generalizability of the network's classification capacity. We explicitly avoided fine-tuning our simulation pipeline to emulate observations of a particular experiment and instead remained agnostic to serve as a proof-of-concept that could be relevant to any survey that produces similar images (i.e. it would not be valid for interferometric data, for example), so this would require matching the experimental specifications of the images in each survey as well as certainly adding complexity to the simulated data. Furthermore, while in this work we fixed the radius of the Einstein ring to the typical value of galaxy-galaxy lenses, and did not add an external shear component, we have to understand whether the network's capacity is robust to varying these two macro-model components, since known galaxy-scale strong lens images span a range of values for both of these parameters and they can be degenerate with subhalo properties.
Another important aspect of the generalizability of the network relates to the implicit modelling of the smooth component of the lens. Although we have not done any explicit smooth modeling to determine the presence or absence of substructure, the network has implicitly learned about the SIE density profile since all the samples in our training data had smooth SIE components. Although many galaxy-scale lenses seem to be well fit with SIE profiles, gauging the network's performance when trained on images simulated with a variety of density profiles would be advantageous before testing on real images. Similarly, training the network on images where the source and lens redshifts vary, the PSF is allowed to be anisotropic, and additional models for the source 
%and macro 
components are included, could also reveal important information about the network's applicability to real data. Understanding all of these factors would be crucial in order to be able to leverage the images classified as non-detections to constrain CDM together with the information extracted from images with substructure. Finally, it would be advantageous to develop a method to quantify confidence in classification, for instance using Gaussian Processes, so that images that are likely to be false positives/negatives can be identified. We leave these to future work.

In addition, we emphasize that there is no reason to believe that an approach such as this one would not be valid for images from an experiment such as ALMA, if the network were trained with an appropriate dataset. This could be an interesting extension of this work, since lens modeling in configuration (uv-visibility) space is even more time consuming than in real space, and experiments such as ALMA are expected to produce very high-quality strong lens images.

The results presented in this work were produced with a CNN whose architecture was the result of a grid search using training/validation samples from the $N_{\rm sub}$-bound dataset in which the image had 1\% uncorrelated noise. It is therefore possible that the architecture could be optimized further for the different datasets, improving these results. More generally, if one had a specific experiment in mind, a grid search could be carried out with images that contained the expected levels of noise and any other relevant experimental details, such as the (possibly anisotropic) PSF. 

Furthermore, the network architecture used in this work is very simple, and has few parameters compared to many typical convolutional net architectures used in the literature. For instance, well-known networks such as AlexNet \cite{imagenet}, GoogLeNet \cite{googlenet}, ResNet \cite{resnet} and DenseNet \cite{densenet} can have tens of millions of parameters. These typically have many more convolutional layers and use additional tricks for training; for example, DenseNet has ``dense blocks", within which the feature maps at each layer are concatenated to the input of every successive layer within the block, allowing later layers to leverage information from earlier layers. 
For instance, our results seem to suggest that the network cannot leverage information from the collective perturbations of lower mass halos, while the recent results from Refs. \cite{mining_substructure} and \cite{stephon}, which both used CNNs based on ResNet to infer different aspects of substructure populations from strong lenses (the subhalo mass function normalization and slope in the former, and to distinguish between substructure pertaining to two very different dark matter scenarios in the latter), show that machine learning methods can in fact be used to probe more than the single most massive halos in lenses. 

We are therefore optimistic that there is room for improvement with respect to the classification capacity of the network we are using here by using a more complex CNN architecture. This is important since, if such an approach were to be pursued moving forward, uncertainties derived on, for example, substructure population parameters or likelihood of CDM, would be inextricably tied to the network's classification accuracy. 

A final remark we want to bring up is with regard to the perturbations from substructure versus from line-of-sight (LOS) halos outside of the main lens halo. Original studies about perturbations to lensed images focused on subhalos as the perturbers. However, it has been pointed out that the contribution of the latter is actually likely to be comparable to, or even greater than, that of the substructure within the lens \cite{Li_los,Despali_los}, meaning that any attempt to use strong lensing images to constrain the particle nature of dark matter must take into account both contributions. This is particularly relevant because the effect of a LOS halo between the observer and the lens is larger than that of a subhalo of the same mass \cite{Despali_los}. What this means is that in fact the sensitivity limits we give here for the \textit{subhalo} mass function substructure can actually translate into sensitivity to lower masses in the full \textit{halo} mass function if we consider LOS halos. 

Strong gravitational lensing as a probe of the particle nature of dark matter has harnessed much interest over the last few years. This, together with the advent of a huge increase in the amount of high-quality strong lens images available for dark matter science, has led to an explosion of research into methods of extracting information from strong lens images. Previous work has used ML to infer strong lens parameters \cite{Hezaveh_cnn1,PerreaultLevasseur_cnn2,Morningstar:2018ase}, to reconstruct the sources from strong lens images \cite{Morningstar:2019szx}, and most recently to infer properties of the substructure population \cite{mining_substructure,stephon}. Now, this work is another step forwards towards understanding the usability of deep learning methods to speed up the analysis of strong lens images for dark matter science. With the considerable momentum that this subfield is gaining, it is possible that in the near future strong lensing will consolidate itself as one of the premier ways to uncover the nature of dark matter. 

\acknowledgements

We thank Francis-Yan Cyr-Racine, Tansu Daylan, Bryan Ostdiek, Simon Birrer and Andrew Robertson for useful comments.
CD was supported by the Department of Energy (DOE) grant DE-SC0020223 and ADR was partially supported by a Dean's Competitive Fund for Promising Scholarship at Harvard University.

\appendix 

\section{Subhalo population characteristics}

Here we show relevant features of the subhalo populations in the images that were simulated following the procedure in Section \ref{sec:data_and_methods}, and are in the training/validation/test sets we used.

\bigskip 

\paragraph{Single subhalo.} 

\begin{figure*}
\includegraphics[width=0.75\textwidth]{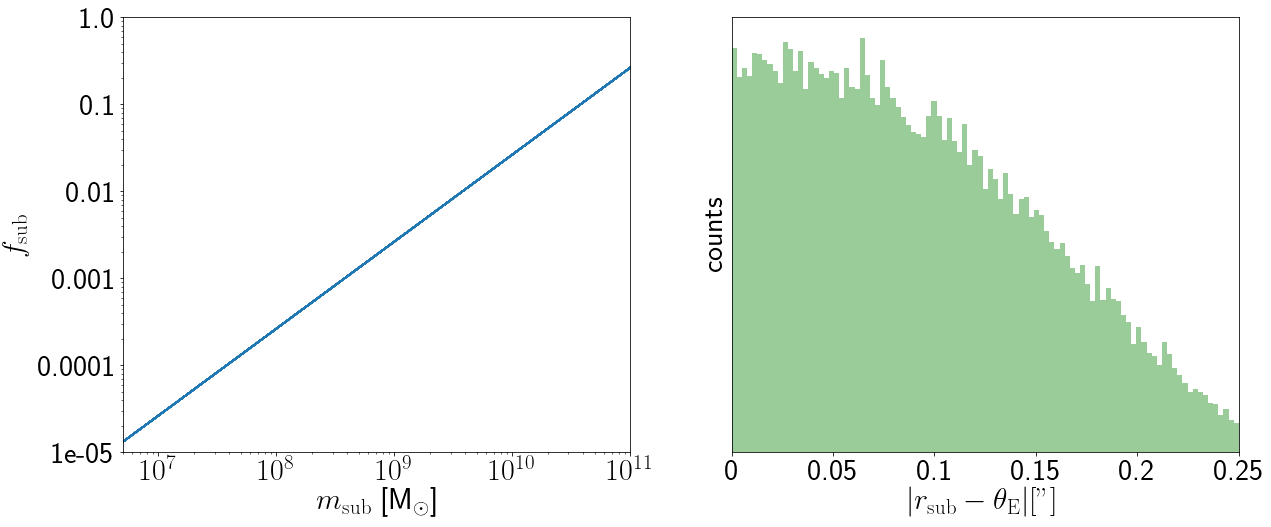}
\caption{\footnotesize{\textbf{Images with a single subhalo} \textit{Left}: mass fraction in substructure $f_{\rm sub}$ as a function of the mass of the subhalo in the image. \textit{Right}: absolute distance from the subhalo's position $r_{\rm sub}$ to the Einstein ring $\theta_{\rm E}$ in arcseconds. }}\label{fig:onesub}
\end{figure*}

Figure \ref{fig:onesub} shows, on the left, the correspondence between the subhalo mass and the mass fraction in substructure, and on the right the absolute distance between the subhalo's position and the Einstein ring. Since these images have a single subhalo, the correspondence between $f_{\rm sub}$ and $m_{\rm sub}$ is trivially one-to-one. Also, notice that the  distribution of the subhalo's position $r_{\rm sub}$ is not uniform from 0" - 0.25" because of the additional constraint on the minimum intensity at the subhalo position. 

\bigskip 

\paragraph{$N_{\rm sub}$-bound.} 

\begin{figure*}
    \centering
    \begin{subfigure}[b]{\textwidth}
        \includegraphics[width=0.75\textwidth]{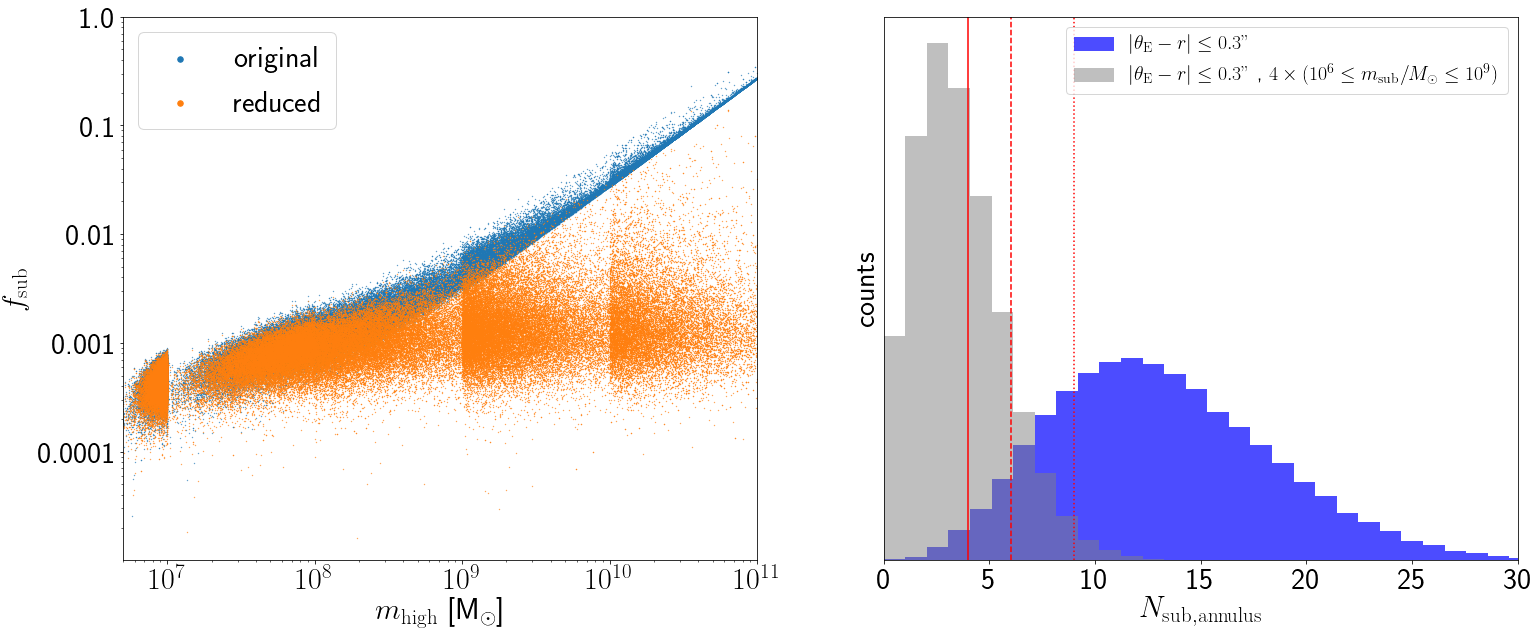}
    \end{subfigure}%
    
    \begin{subfigure}[b]{0.6\textwidth}
        \includegraphics[width=0.6\textwidth]{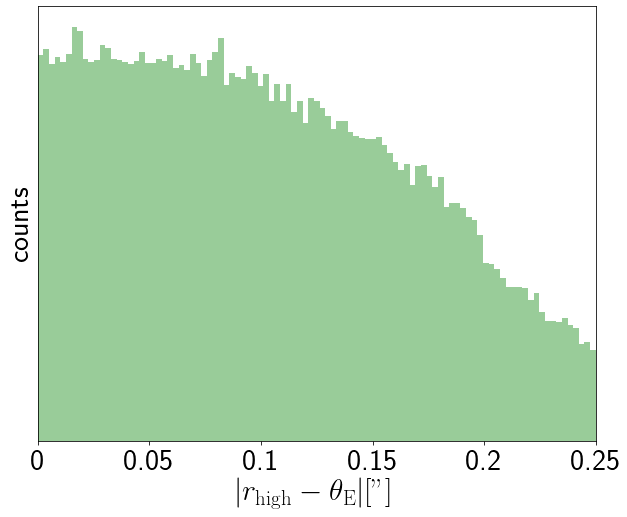}
    \end{subfigure}

\caption{\footnotesize{
\textbf{Images with $N_{\rm sub}$ constrained} \textit{Top left}: in blue, the mass fraction in substructure $f_{\rm sub}$ as a function of the highest subhalo mass in the image $m_{\rm high}$. The ``reduced" markers correspond to having removed the most massive subhalo. 
\textit{Top right}: in blue, the total number of subhalos in an annulus 0.6'' wide centered on the Einstein ring. In gray, the number of subhalos with masses between $4 \times 10^6 - 4\times 10^9$ M$_{\odot}$ in this same area. The solid, dashed, and dotted red lines correspond to the 50th, 16th/84th and 2.5th/97.5th percentiles, respectively. \textit{Bottom}: absolute distance from the position of the most massive subhalo $r_{\rm high}$ to the Einstein ring $\theta_{\rm E}$, in units of arcseconds.}}\label{fig:nsub}
\end{figure*}

Figure \ref{fig:nsub} shows different features of the subhalo populations over the entire training/validation/test sets images, where the number of subhalos per image was constrained. The top left panel shows the fraction of mass in substructure as a function of the highest-mass subhalo in the image. Most images have a fraction of mass in substructure well below 1\% (in particular all the images with $m_{\rm high} < 10^9$ M$_{\rm \odot}$) and, as expected, $f_{\rm sub}$ increases considerably with increasing $m_{\rm high}$. The orange markers show the value of $f_{\rm sub}$ if the most massive subhalo in the image is removed. This shows that for the majority of the images with $m_{\rm high} \gtrsim 10^9$ M$_{\odot}$ the value of $f_{\rm sub}$ is driven by the single most massive halo, although there is some scatter. In Section \ref{sec:results} we discuss the effect (or lack thereof) of the images with high $f_{\rm sub}$ due to more than a single subhalo.

The top right panel shows the number of subhalos within an annulus of width 0.6" centered on the Einstein ring together with a histogram of the number of subhalos in the annulus only with masses between $4 \times 10^6 - 4 \times 10^9$ M$_{\odot}$, to have a direct point of comparison to the numbers cited in \cite{2009MNRAS.400.1583V}, where it was found that, in the context of CDM, for $f_{\rm sub} = 0.3\%$ there should be $\sim 7 \pm 1$ subhalos with masses between $4 \times 10^6 - 4 \times 10^9$ M$_{\odot}$ in an annulus of this width. We can see that the number of subhalos (and $f_{\rm sub}$) lie comfortably in the lower end of the expectations within CDM. Finally, the bottom panel shows the absolute distance between the position of the most massive subhalo and the Einstein radius. Again it is apparent that the distribution is not perfectly uniform due to the fact that the most massive subhalo has to lie at a point on the image with non-negligible intensity. 

\bigskip 

\paragraph{$f_{\rm sub}$-bound.} 

\begin{figure*}
    \centering
    \begin{subfigure}[b]{\textwidth}
        \includegraphics[width=0.75\textwidth]{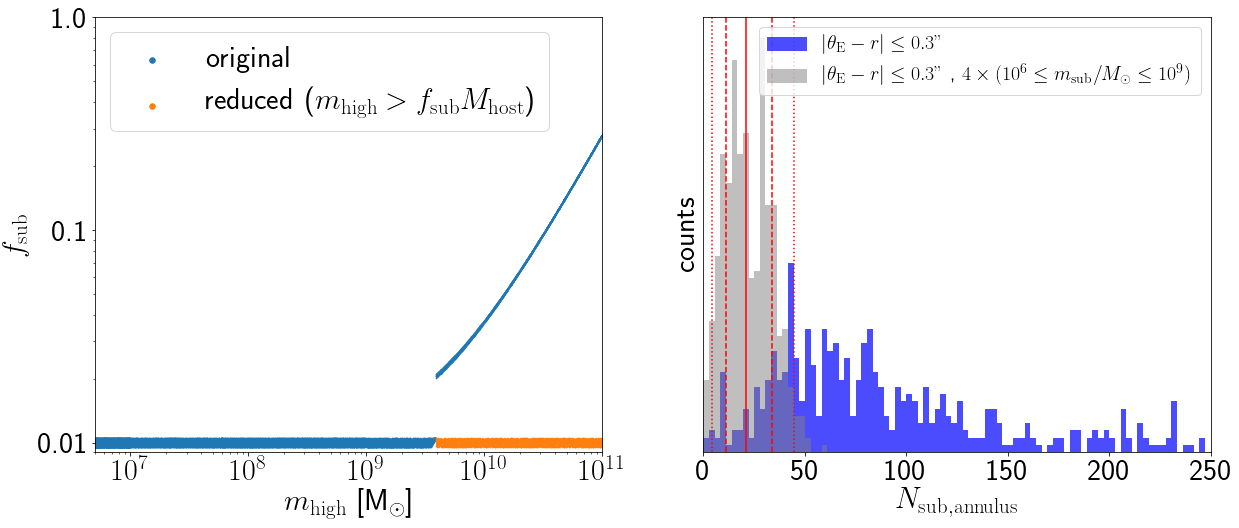}
    \end{subfigure}%
    
    \begin{subfigure}[b]{0.6\textwidth}
        \includegraphics[width=0.6\textwidth]{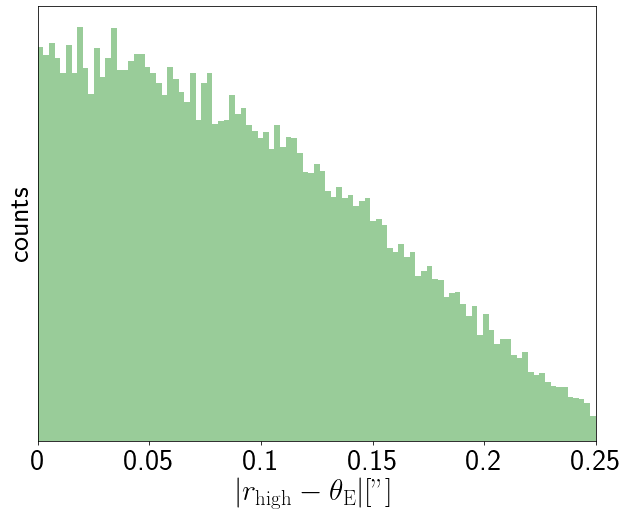}
    \end{subfigure}
    
\caption{\footnotesize{
\textbf{Images with $f_{\rm sub}$ constrained} \textit{Top left}: in blue, the mass fraction in substructure $f_{\rm sub}$ as a function of the highest subhalo mass in the image $m_{\rm high}$. The ``reduced" markers correspond to having removed the most massive subhalo for images with $m_{\rm high} > f_{\rm sub} M_{\rm host}$ to show that for these images $f_{\rm sub}$ is still bound to $1\pm0.05\%$ when the most massive subhalo isn't taken into account. 
\textit{Top right}: in blue, the total number of subhalos in an annulus 0.6'' wide centered on the Einstein ring. In gray, the number of subhalos with masses between $4 \times 10^6 - 4\times 10^9$ M$_{\odot}$ in this same area. The solid, dashed, and dotted red lines correspond to the 50th, 16th/84th and 2.5th/97.5th percentiles, respectively. \textit{Bottom}: absolute distance from the position of the most massive subhalo $r_{\rm high}$ to the Einstein ring $\theta_{\rm E}$, in units of arcseconds.}}\label{fig:fsub}
\end{figure*}

Figure \ref{fig:fsub} is analogous to Figure \ref{fig:nsub} but for the dataset where the mass fraction in substructure per image was constrained to be $f_{\rm sub} = 1 \pm 0.05\%$. The top left panel shows that, up to $f_{\rm sub} M_{\rm host}$, the mass fraction in substructure is $1 \pm 0.05 \%$. Past this point, as explained in Section \ref{sec:contrain_fsub}, the single most massive subhalo already saturates this bound so instead, for a given $m_{\rm high} > f_{\rm sub} M_{\rm host}$, we generate a population of subhalos that does obey the bound on $f_{\rm sub}$ but has $m_{\rm high}' < m_{\rm high}$, and then add $m_{\rm high}$ to the image. The orange markers show the ``reduced" value of $f_{\rm sub}$, where the most massive subhalo isn't taken into account. It can be seen that it obeys the constraint of $1\pm0.05\%$. 

The top right panel again shows the number of subhalos in an annulus that is 0.6" wide centered on the Einstein ring in blue, and in gray the number of subhalos with masses between $4 \times 10^6 - 4 \times 10^9$ M$_{\odot}$. It has a mean of 21 subhalos, which is expected since these images have a $\sim 3\times$ greater value of $f_{\rm sub}$ than those in Ref. \cite{2009MNRAS.400.1583V}. The bottom panel shows, as before, the absolute distance from the position of the most massive subhalo to the Einstein ring.

\newpage

\bibliography{main}

\begin{thebibliography}{71}%
\makeatletter
\providecommand \@ifxundefined [1]{%
 \@ifx{#1\undefined}
}%
\providecommand \@ifnum [1]{%
 \ifnum #1\expandafter \@firstoftwo
 \else \expandafter \@secondoftwo
 \fi
}%
\providecommand \@ifx [1]{%
 \ifx #1\expandafter \@firstoftwo
 \else \expandafter \@secondoftwo
 \fi
}%
\providecommand \natexlab [1]{#1}%
\providecommand \enquote  [1]{``#1''}%
\providecommand \bibnamefont  [1]{#1}%
\providecommand \bibfnamefont [1]{#1}%
\providecommand \citenamefont [1]{#1}%
\providecommand \href@noop [0]{\@secondoftwo}%
\providecommand \href [0]{\begingroup \@sanitize@url \@href}%
\providecommand \@href[1]{\@@startlink{#1}\@@href}%
\providecommand \@@href[1]{\endgroup#1\@@endlink}%
\providecommand \@sanitize@url [0]{\catcode `\\12\catcode `\$12\catcode
  `\&12\catcode `\#12\catcode `\^12\catcode `\_12\catcode `\%12\relax}%
\providecommand \@@startlink[1]{}%
\providecommand \@@endlink[0]{}%
\providecommand \url  [0]{\begingroup\@sanitize@url \@url }%
\providecommand \@url [1]{\endgroup\@href {#1}{\urlprefix }}%
\providecommand \urlprefix  [0]{URL }%
\providecommand \Eprint [0]{\href }%
\providecommand \doibase [0]{http://dx.doi.org/}%
\providecommand \selectlanguage [0]{\@gobble}%
\providecommand \bibinfo  [0]{\@secondoftwo}%
\providecommand \bibfield  [0]{\@secondoftwo}%
\providecommand \translation [1]{[#1]}%
\providecommand \BibitemOpen [0]{}%
\providecommand \bibitemStop [0]{}%
\providecommand \bibitemNoStop [0]{.\EOS\space}%
\providecommand \EOS [0]{\spacefactor3000\relax}%
\providecommand \BibitemShut  [1]{\csname bibitem#1\endcsname}%
\let\auto@bib@innerbib\@empty
%</preamble>
\bibitem [{\citenamefont {Alam}\ \emph {et~al.}(2017)\citenamefont {Alam} \emph
  {et~al.}}]{BOSS}%
  \BibitemOpen
  \bibfield  {author} {\bibinfo {author} {\bibfnamefont {S.}~\bibnamefont
  {Alam}} \emph {et~al.} (\bibinfo {collaboration} {BOSS}),\ }\href {\doibase
  10.1093/mnras/stx721} {\bibfield  {journal} {\bibinfo  {journal} {Mon. Not.
  Roy. Astron. Soc.}\ }\textbf {\bibinfo {volume} {470}},\ \bibinfo {pages}
  {2617} (\bibinfo {year} {2017})},\ \Eprint {http://arxiv.org/abs/1607.03155}
  {arXiv:1607.03155 [astro-ph.CO]} \BibitemShut {NoStop}%
%%CITATION = ARXIV:1607.03155;%%
\bibitem [{\citenamefont {Scolnic}\ \emph {et~al.}(2018)\citenamefont {Scolnic}
  \emph {et~al.}}]{Pantheon}%
  \BibitemOpen
  \bibfield  {author} {\bibinfo {author} {\bibfnamefont {D.~M.}\ \bibnamefont
  {Scolnic}} \emph {et~al.},\ }\href {\doibase 10.3847/1538-4357/aab9bb}
  {\bibfield  {journal} {\bibinfo  {journal} {Astrophys. J.}\ }\textbf
  {\bibinfo {volume} {859}},\ \bibinfo {pages} {101} (\bibinfo {year}
  {2018})},\ \Eprint {http://arxiv.org/abs/1710.00845} {arXiv:1710.00845
  [astro-ph.CO]} \BibitemShut {NoStop}%
%%CITATION = ARXIV:1710.00845;%%
\bibitem [{\citenamefont {Aghanim}\ \emph {et~al.}(2018)\citenamefont {Aghanim}
  \emph {et~al.}}]{Planck2018}%
  \BibitemOpen
  \bibfield  {author} {\bibinfo {author} {\bibfnamefont {N.}~\bibnamefont
  {Aghanim}} \emph {et~al.} (\bibinfo {collaboration} {Planck}),\ }\href@noop
  {} {\  (\bibinfo {year} {2018})},\ \Eprint {http://arxiv.org/abs/1807.06209}
  {arXiv:1807.06209 [astro-ph.CO]} \BibitemShut {NoStop}%
%%CITATION = ARXIV:1807.06209;%%
\bibitem [{\citenamefont {Troxel}\ \emph {et~al.}(2018)\citenamefont {Troxel}
  \emph {et~al.}}]{S8_tension}%
  \BibitemOpen
  \bibfield  {author} {\bibinfo {author} {\bibfnamefont {M.~A.}\ \bibnamefont
  {Troxel}} \emph {et~al.} (\bibinfo {collaboration} {DES}),\ }\href {\doibase
  10.1103/PhysRevD.98.043528} {\bibfield  {journal} {\bibinfo  {journal} {Phys.
  Rev.}\ }\textbf {\bibinfo {volume} {D98}},\ \bibinfo {pages} {043528}
  (\bibinfo {year} {2018})},\ \Eprint {http://arxiv.org/abs/1708.01538}
  {arXiv:1708.01538 [astro-ph.CO]} \BibitemShut {NoStop}%
%%CITATION = ARXIV:1708.01538;%%
\bibitem [{\citenamefont {Verde}\ \emph {et~al.}(2019)\citenamefont {Verde},
  \citenamefont {Treu},\ and\ \citenamefont {Riess}}]{Tensions}%
  \BibitemOpen
  \bibfield  {author} {\bibinfo {author} {\bibfnamefont {L.}~\bibnamefont
  {Verde}}, \bibinfo {author} {\bibfnamefont {T.}~\bibnamefont {Treu}}, \ and\
  \bibinfo {author} {\bibfnamefont {A.~G.}\ \bibnamefont {Riess}}\ }(\bibinfo
  {year} {2019})\ \Eprint {http://arxiv.org/abs/1907.10625} {arXiv:1907.10625
  [astro-ph.CO]} \BibitemShut {NoStop}%
%%CITATION = ARXIV:1907.10625;%%
\bibitem [{\citenamefont {{Kravtsov}}(2010)}]{missing_satellites}%
  \BibitemOpen
  \bibfield  {author} {\bibinfo {author} {\bibfnamefont {A.}~\bibnamefont
  {{Kravtsov}}},\ }\href {\doibase 10.1155/2010/281913} {\bibfield  {journal}
  {\bibinfo  {journal} {Advances in Astronomy}\ }\textbf {\bibinfo {volume}
  {2010}},\ \bibinfo {eid} {281913} (\bibinfo {year} {2010})},\ \Eprint
  {http://arxiv.org/abs/0906.3295} {arXiv:0906.3295 [astro-ph.CO]} \BibitemShut
  {NoStop}%
\bibitem [{\citenamefont {{Boylan-Kolchin}}\ \emph {et~al.}(2011)\citenamefont
  {{Boylan-Kolchin}}, \citenamefont {{Bullock}},\ and\ \citenamefont
  {{Kaplinghat}}}]{TBTF1}%
  \BibitemOpen
  \bibfield  {author} {\bibinfo {author} {\bibfnamefont {M.}~\bibnamefont
  {{Boylan-Kolchin}}}, \bibinfo {author} {\bibfnamefont {J.~S.}\ \bibnamefont
  {{Bullock}}}, \ and\ \bibinfo {author} {\bibfnamefont {M.}~\bibnamefont
  {{Kaplinghat}}},\ }\href {\doibase 10.1111/j.1745-3933.2011.01074.x}
  {\bibfield  {journal} {\bibinfo  {journal} {\mnras}\ }\textbf {\bibinfo
  {volume} {415}},\ \bibinfo {pages} {L40} (\bibinfo {year} {2011})},\ \Eprint
  {http://arxiv.org/abs/1103.0007} {arXiv:1103.0007 [astro-ph.CO]} \BibitemShut
  {NoStop}%
\bibitem [{\citenamefont {{Boylan-Kolchin}}\ \emph {et~al.}(2012)\citenamefont
  {{Boylan-Kolchin}}, \citenamefont {{Bullock}},\ and\ \citenamefont
  {{Kaplinghat}}}]{TBTF2}%
  \BibitemOpen
  \bibfield  {author} {\bibinfo {author} {\bibfnamefont {M.}~\bibnamefont
  {{Boylan-Kolchin}}}, \bibinfo {author} {\bibfnamefont {J.~S.}\ \bibnamefont
  {{Bullock}}}, \ and\ \bibinfo {author} {\bibfnamefont {M.}~\bibnamefont
  {{Kaplinghat}}},\ }\href {\doibase 10.1111/j.1365-2966.2012.20695.x}
  {\bibfield  {journal} {\bibinfo  {journal} {\mnras}\ }\textbf {\bibinfo
  {volume} {422}},\ \bibinfo {pages} {1203} (\bibinfo {year} {2012})},\ \Eprint
  {http://arxiv.org/abs/1111.2048} {arXiv:1111.2048 [astro-ph.CO]} \BibitemShut
  {NoStop}%
\bibitem [{\citenamefont {Ngan}\ and\ \citenamefont
  {Carlberg}(2014)}]{Ngan:2013oga}%
  \BibitemOpen
  \bibfield  {author} {\bibinfo {author} {\bibfnamefont {W.-H.~W.}\
  \bibnamefont {Ngan}}\ and\ \bibinfo {author} {\bibfnamefont {R.~G.}\
  \bibnamefont {Carlberg}},\ }\href {\doibase 10.1088/0004-637X/788/2/181}
  {\bibfield  {journal} {\bibinfo  {journal} {Astrophys. J.}\ }\textbf
  {\bibinfo {volume} {788}},\ \bibinfo {pages} {181} (\bibinfo {year}
  {2014})},\ \Eprint {http://arxiv.org/abs/1311.1710} {arXiv:1311.1710
  [astro-ph.CO]} \BibitemShut {NoStop}%
%%CITATION = ARXIV:1311.1710;%%
\bibitem [{\citenamefont {{Carlberg}}(2016)}]{2016ApJ...820...45C}%
  \BibitemOpen
  \bibfield  {author} {\bibinfo {author} {\bibfnamefont {R.~G.}\ \bibnamefont
  {{Carlberg}}},\ }\href {\doibase 10.3847/0004-637X/820/1/45} {\bibfield
  {journal} {\bibinfo  {journal} {\apj}\ }\textbf {\bibinfo {volume} {820}},\
  \bibinfo {eid} {45} (\bibinfo {year} {2016})},\ \Eprint
  {http://arxiv.org/abs/1512.01620} {arXiv:1512.01620 [astro-ph.GA]}
  \BibitemShut {NoStop}%
\bibitem [{\citenamefont {Bovy}(2016)}]{Bovy:2015mda}%
  \BibitemOpen
  \bibfield  {author} {\bibinfo {author} {\bibfnamefont {J.}~\bibnamefont
  {Bovy}},\ }\href {\doibase 10.1103/PhysRevLett.116.121301} {\bibfield
  {journal} {\bibinfo  {journal} {Phys. Rev. Lett.}\ }\textbf {\bibinfo
  {volume} {116}},\ \bibinfo {pages} {121301} (\bibinfo {year} {2016})},\
  \Eprint {http://arxiv.org/abs/1512.00452} {arXiv:1512.00452 [astro-ph.GA]}
  \BibitemShut {NoStop}%
%%CITATION = ARXIV:1512.00452;%%
\bibitem [{\citenamefont {{Erkal}}\ \emph {et~al.}(2016)\citenamefont
  {{Erkal}}, \citenamefont {{Belokurov}}, \citenamefont {{Bovy}},\ and\
  \citenamefont {{Sand ers}}}]{2016MNRAS.463..102E}%
  \BibitemOpen
  \bibfield  {author} {\bibinfo {author} {\bibfnamefont {D.}~\bibnamefont
  {{Erkal}}}, \bibinfo {author} {\bibfnamefont {V.}~\bibnamefont
  {{Belokurov}}}, \bibinfo {author} {\bibfnamefont {J.}~\bibnamefont {{Bovy}}},
  \ and\ \bibinfo {author} {\bibfnamefont {J.~L.}\ \bibnamefont {{Sand ers}}},\
  }\href {\doibase 10.1093/mnras/stw1957} {\bibfield  {journal} {\bibinfo
  {journal} {\mnras}\ }\textbf {\bibinfo {volume} {463}},\ \bibinfo {pages}
  {102} (\bibinfo {year} {2016})},\ \Eprint {http://arxiv.org/abs/1606.04946}
  {arXiv:1606.04946 [astro-ph.GA]} \BibitemShut {NoStop}%
\bibitem [{\citenamefont {Feldmann}\ and\ \citenamefont
  {Spolyar}(2015)}]{Feldmann:2013hqa}%
  \BibitemOpen
  \bibfield  {author} {\bibinfo {author} {\bibfnamefont {R.}~\bibnamefont
  {Feldmann}}\ and\ \bibinfo {author} {\bibfnamefont {D.}~\bibnamefont
  {Spolyar}},\ }\href {\doibase 10.1093/mnras/stu2147} {\bibfield  {journal}
  {\bibinfo  {journal} {Mon. Not. Roy. Astron. Soc.}\ }\textbf {\bibinfo
  {volume} {446}},\ \bibinfo {pages} {1000} (\bibinfo {year} {2015})},\ \Eprint
  {http://arxiv.org/abs/1310.2243} {arXiv:1310.2243 [astro-ph.GA]} \BibitemShut
  {NoStop}%
%%CITATION = ARXIV:1310.2243;%%
\bibitem [{\citenamefont {Mao}\ and\ \citenamefont {Schneider}(1998)}]{Mao}%
  \BibitemOpen
  \bibfield  {author} {\bibinfo {author} {\bibfnamefont {S.-d.}\ \bibnamefont
  {Mao}}\ and\ \bibinfo {author} {\bibfnamefont {P.}~\bibnamefont
  {Schneider}},\ }\href {\doibase 10.1046/j.1365-8711.1998.01319.x} {\bibfield
  {journal} {\bibinfo  {journal} {Mon. Not. Roy. Astron. Soc.}\ }\textbf
  {\bibinfo {volume} {295}},\ \bibinfo {pages} {587} (\bibinfo {year}
  {1998})},\ \Eprint {http://arxiv.org/abs/astro-ph/9707187}
  {arXiv:astro-ph/9707187 [astro-ph]} \BibitemShut {NoStop}%
%%CITATION = ASTRO-PH/9707187;%%
\bibitem [{\citenamefont {Koopmans}(2005)}]{grav_imaging1}%
  \BibitemOpen
  \bibfield  {author} {\bibinfo {author} {\bibfnamefont {L.~V.~E.}\
  \bibnamefont {Koopmans}},\ }\href {\doibase 10.1111/j.1365-2966.2005.09523.x}
  {\bibfield  {journal} {\bibinfo  {journal} {Mon. Not. Roy. Astron. Soc.}\
  }\textbf {\bibinfo {volume} {363}},\ \bibinfo {pages} {1136} (\bibinfo {year}
  {2005})},\ \Eprint {http://arxiv.org/abs/astro-ph/0501324}
  {arXiv:astro-ph/0501324 [astro-ph]} \BibitemShut {NoStop}%
%%CITATION = ASTRO-PH/0501324;%%
\bibitem [{\citenamefont {Vegetti}\ and\ \citenamefont
  {Koopmans}(2009)}]{grav_imaging2}%
  \BibitemOpen
  \bibfield  {author} {\bibinfo {author} {\bibfnamefont {S.}~\bibnamefont
  {Vegetti}}\ and\ \bibinfo {author} {\bibfnamefont {L.~V.~E.}\ \bibnamefont
  {Koopmans}},\ }\href {\doibase 10.1111/j.1365-2966.2008.14005.x} {\bibfield
  {journal} {\bibinfo  {journal} {Mon. Not. Roy. Astron. Soc.}\ }\textbf
  {\bibinfo {volume} {392}},\ \bibinfo {pages} {945} (\bibinfo {year}
  {2009})},\ \Eprint {http://arxiv.org/abs/0805.0201} {arXiv:0805.0201
  [astro-ph]} \BibitemShut {NoStop}%
%%CITATION = ARXIV:0805.0201;%%
\bibitem [{\citenamefont {Moustakas}\ and\ \citenamefont
  {Metcalf}(2003)}]{spat_res_spec1}%
  \BibitemOpen
  \bibfield  {author} {\bibinfo {author} {\bibfnamefont {L.~A.}\ \bibnamefont
  {Moustakas}}\ and\ \bibinfo {author} {\bibfnamefont {R.~B.}\ \bibnamefont
  {Metcalf}},\ }\href {\doibase 10.1046/j.1365-8711.2003.06055.x} {\bibfield
  {journal} {\bibinfo  {journal} {Mon. Not. Roy. Astron. Soc.}\ }\textbf
  {\bibinfo {volume} {339}},\ \bibinfo {pages} {607} (\bibinfo {year}
  {2003})},\ \Eprint {http://arxiv.org/abs/astro-ph/0206176}
  {arXiv:astro-ph/0206176 [astro-ph]} \BibitemShut {NoStop}%
%%CITATION = ASTRO-PH/0206176;%%
\bibitem [{\citenamefont {{Hezaveh}}\ \emph {et~al.}(2013)\citenamefont
  {{Hezaveh}}, \citenamefont {{Dalal}}, \citenamefont {{Holder}}, \citenamefont
  {{Kuhlen}}, \citenamefont {{Marrone}}, \citenamefont {{Murray}},\ and\
  \citenamefont {{Vieira}}}]{spat_res_spec2}%
  \BibitemOpen
  \bibfield  {author} {\bibinfo {author} {\bibfnamefont {Y.}~\bibnamefont
  {{Hezaveh}}}, \bibinfo {author} {\bibfnamefont {N.}~\bibnamefont {{Dalal}}},
  \bibinfo {author} {\bibfnamefont {G.}~\bibnamefont {{Holder}}}, \bibinfo
  {author} {\bibfnamefont {M.}~\bibnamefont {{Kuhlen}}}, \bibinfo {author}
  {\bibfnamefont {D.}~\bibnamefont {{Marrone}}}, \bibinfo {author}
  {\bibfnamefont {N.}~\bibnamefont {{Murray}}}, \ and\ \bibinfo {author}
  {\bibfnamefont {J.}~\bibnamefont {{Vieira}}},\ }\href {\doibase
  10.1088/0004-637X/767/1/9} {\bibfield  {journal} {\bibinfo  {journal} {\apj}\
  }\textbf {\bibinfo {volume} {767}},\ \bibinfo {eid} {9} (\bibinfo {year}
  {2013})},\ \Eprint {http://arxiv.org/abs/1210.4562} {arXiv:1210.4562
  [astro-ph.CO]} \BibitemShut {NoStop}%
\bibitem [{\citenamefont {Hezaveh}\ \emph {et~al.}(2016)\citenamefont
  {Hezaveh}, \citenamefont {Dalal}, \citenamefont {Holder}, \citenamefont
  {Kisner}, \citenamefont {Kuhlen},\ and\ \citenamefont
  {Perreault~Levasseur}}]{Hezaveh_powerspec}%
  \BibitemOpen
  \bibfield  {author} {\bibinfo {author} {\bibfnamefont {Y.}~\bibnamefont
  {Hezaveh}}, \bibinfo {author} {\bibfnamefont {N.}~\bibnamefont {Dalal}},
  \bibinfo {author} {\bibfnamefont {G.}~\bibnamefont {Holder}}, \bibinfo
  {author} {\bibfnamefont {T.}~\bibnamefont {Kisner}}, \bibinfo {author}
  {\bibfnamefont {M.}~\bibnamefont {Kuhlen}}, \ and\ \bibinfo {author}
  {\bibfnamefont {L.}~\bibnamefont {Perreault~Levasseur}},\ }\href {\doibase
  10.1088/1475-7516/2016/11/048} {\bibfield  {journal} {\bibinfo  {journal}
  {JCAP}\ }\textbf {\bibinfo {volume} {1611}},\ \bibinfo {pages} {048}
  (\bibinfo {year} {2016})},\ \Eprint {http://arxiv.org/abs/1403.2720}
  {arXiv:1403.2720 [astro-ph.CO]} \BibitemShut {NoStop}%
%%CITATION = ARXIV:1403.2720;%%
\bibitem [{\citenamefont {Daylan}\ \emph {et~al.}(2018)\citenamefont {Daylan},
  \citenamefont {Cyr-Racine}, \citenamefont {Diaz~Rivero}, \citenamefont
  {Dvorkin},\ and\ \citenamefont {Finkbeiner}}]{pcatlens}%
  \BibitemOpen
  \bibfield  {author} {\bibinfo {author} {\bibfnamefont {T.}~\bibnamefont
  {Daylan}}, \bibinfo {author} {\bibfnamefont {F.-Y.}\ \bibnamefont
  {Cyr-Racine}}, \bibinfo {author} {\bibfnamefont {A.}~\bibnamefont
  {Diaz~Rivero}}, \bibinfo {author} {\bibfnamefont {C.}~\bibnamefont
  {Dvorkin}}, \ and\ \bibinfo {author} {\bibfnamefont {D.~P.}\ \bibnamefont
  {Finkbeiner}},\ }\href {\doibase 10.3847/1538-4357/aaaa1e} {\bibfield
  {journal} {\bibinfo  {journal} {Astrophys. J.}\ }\textbf {\bibinfo {volume}
  {854}},\ \bibinfo {pages} {141} (\bibinfo {year} {2018})},\ \Eprint
  {http://arxiv.org/abs/1706.06111} {arXiv:1706.06111 [astro-ph.CO]}
  \BibitemShut {NoStop}%
%%CITATION = ARXIV:1706.06111;%%
\bibitem [{\citenamefont {Birrer}\ \emph {et~al.}(2017)\citenamefont {Birrer},
  \citenamefont {Amara},\ and\ \citenamefont {Refregier}}]{Birrer:2017rpp}%
  \BibitemOpen
  \bibfield  {author} {\bibinfo {author} {\bibfnamefont {S.}~\bibnamefont
  {Birrer}}, \bibinfo {author} {\bibfnamefont {A.}~\bibnamefont {Amara}}, \
  and\ \bibinfo {author} {\bibfnamefont {A.}~\bibnamefont {Refregier}},\ }\href
  {\doibase 10.1088/1475-7516/2017/05/037} {\bibfield  {journal} {\bibinfo
  {journal} {JCAP}\ }\textbf {\bibinfo {volume} {1705}},\ \bibinfo {pages}
  {037} (\bibinfo {year} {2017})},\ \Eprint {http://arxiv.org/abs/1702.00009}
  {arXiv:1702.00009 [astro-ph.CO]} \BibitemShut {NoStop}%
%%CITATION = ARXIV:1702.00009;%%
\bibitem [{\citenamefont {Brewer}\ \emph {et~al.}(2016)\citenamefont {Brewer},
  \citenamefont {Huijser},\ and\ \citenamefont {Lewis}}]{Brewer:2015yya}%
  \BibitemOpen
  \bibfield  {author} {\bibinfo {author} {\bibfnamefont {B.~J.}\ \bibnamefont
  {Brewer}}, \bibinfo {author} {\bibfnamefont {D.}~\bibnamefont {Huijser}}, \
  and\ \bibinfo {author} {\bibfnamefont {G.~F.}\ \bibnamefont {Lewis}},\ }\href
  {\doibase 10.1093/mnras/stv2370} {\bibfield  {journal} {\bibinfo  {journal}
  {Mon. Not. Roy. Astron. Soc.}\ }\textbf {\bibinfo {volume} {455}},\ \bibinfo
  {pages} {1819} (\bibinfo {year} {2016})},\ \Eprint
  {http://arxiv.org/abs/1508.00662} {arXiv:1508.00662 [astro-ph.IM]}
  \BibitemShut {NoStop}%
%%CITATION = ARXIV:1508.00662;%%
\bibitem [{\citenamefont {Cyr-Racine}\ \emph {et~al.}(2016)\citenamefont
  {Cyr-Racine}, \citenamefont {Moustakas}, \citenamefont {Keeton},
  \citenamefont {Sigurdson},\ and\ \citenamefont {Gilman}}]{dark_census}%
  \BibitemOpen
  \bibfield  {author} {\bibinfo {author} {\bibfnamefont {F.}~\bibnamefont
  {Cyr-Racine}}, \bibinfo {author} {\bibfnamefont {L.}~\bibnamefont
  {Moustakas}}, \bibinfo {author} {\bibfnamefont {C.}~\bibnamefont {Keeton}},
  \bibinfo {author} {\bibfnamefont {K.}~\bibnamefont {Sigurdson}}, \ and\
  \bibinfo {author} {\bibfnamefont {D.}~\bibnamefont {Gilman}},\ }\href
  {\doibase 10.1103/PhysRevD.94.043505} {\bibfield  {journal} {\bibinfo
  {journal} {Physical Review D}\ }\textbf {\bibinfo {volume} {94}} (\bibinfo
  {year} {2016}),\ 10.1103/PhysRevD.94.043505}\BibitemShut {NoStop}%
\bibitem [{\citenamefont {Diaz~Rivero}\ \emph {et~al.}(2018)\citenamefont
  {Diaz~Rivero}, \citenamefont {Cyr-Racine},\ and\ \citenamefont
  {Dvorkin}}]{powerspec1}%
  \BibitemOpen
  \bibfield  {author} {\bibinfo {author} {\bibfnamefont {A.}~\bibnamefont
  {Diaz~Rivero}}, \bibinfo {author} {\bibfnamefont {F.-Y.}\ \bibnamefont
  {Cyr-Racine}}, \ and\ \bibinfo {author} {\bibfnamefont {C.}~\bibnamefont
  {Dvorkin}},\ }\href {\doibase 10.1103/PhysRevD.97.023001} {\bibfield
  {journal} {\bibinfo  {journal} {Phys. Rev.}\ }\textbf {\bibinfo {volume}
  {D97}},\ \bibinfo {pages} {023001} (\bibinfo {year} {2018})},\ \Eprint
  {http://arxiv.org/abs/1707.04590} {arXiv:1707.04590 [astro-ph.CO]}
  \BibitemShut {NoStop}%
%%CITATION = ARXIV:1707.04590;%%
\bibitem [{\citenamefont {Díaz~Rivero}\ \emph {et~al.}(2018)\citenamefont
  {Díaz~Rivero}, \citenamefont {Dvorkin}, \citenamefont {Cyr-Racine},
  \citenamefont {Zavala},\ and\ \citenamefont {Vogelsberger}}]{powerspec2}%
  \BibitemOpen
  \bibfield  {author} {\bibinfo {author} {\bibfnamefont {A.}~\bibnamefont
  {Díaz~Rivero}}, \bibinfo {author} {\bibfnamefont {C.}~\bibnamefont
  {Dvorkin}}, \bibinfo {author} {\bibfnamefont {F.-Y.}\ \bibnamefont
  {Cyr-Racine}}, \bibinfo {author} {\bibfnamefont {J.}~\bibnamefont {Zavala}},
  \ and\ \bibinfo {author} {\bibfnamefont {M.}~\bibnamefont {Vogelsberger}},\
  }\href {\doibase 10.1103/PhysRevD.98.103517} {\bibfield  {journal} {\bibinfo
  {journal} {Phys. Rev.}\ }\textbf {\bibinfo {volume} {D98}},\ \bibinfo {pages}
  {103517} (\bibinfo {year} {2018})},\ \Eprint
  {http://arxiv.org/abs/1809.00004} {arXiv:1809.00004 [astro-ph.CO]}
  \BibitemShut {NoStop}%
%%CITATION = ARXIV:1809.00004;%%
\bibitem [{\citenamefont {Brennan}\ \emph {et~al.}(2019)\citenamefont
  {Brennan}, \citenamefont {Benson}, \citenamefont {Cyr-Racine}, \citenamefont
  {Keeton}, \citenamefont {Moustakas},\ and\ \citenamefont {Pullen}}]{Brennan}%
  \BibitemOpen
  \bibfield  {author} {\bibinfo {author} {\bibfnamefont {S.}~\bibnamefont
  {Brennan}}, \bibinfo {author} {\bibfnamefont {A.~J.}\ \bibnamefont {Benson}},
  \bibinfo {author} {\bibfnamefont {F.-Y.}\ \bibnamefont {Cyr-Racine}},
  \bibinfo {author} {\bibfnamefont {C.~R.}\ \bibnamefont {Keeton}}, \bibinfo
  {author} {\bibfnamefont {L.~A.}\ \bibnamefont {Moustakas}}, \ and\ \bibinfo
  {author} {\bibfnamefont {A.~R.}\ \bibnamefont {Pullen}},\ }\href {\doibase
  10.1093/mnras/stz1607} {\bibfield  {journal} {\bibinfo  {journal} {Mon. Not.
  Roy. Astron. Soc.}\ }\textbf {\bibinfo {volume} {488}},\ \bibinfo {pages}
  {5085} (\bibinfo {year} {2019})},\ \Eprint {http://arxiv.org/abs/1808.03501}
  {arXiv:1808.03501 [astro-ph.GA]} \BibitemShut {NoStop}%
%%CITATION = ARXIV:1808.03501;%%
\bibitem [{\citenamefont {Brehmer}\ \emph {et~al.}(2019)\citenamefont
  {Brehmer}, \citenamefont {Mishra-Sharma}, \citenamefont {Hermans},
  \citenamefont {Louppe},\ and\ \citenamefont {Cranmer}}]{mining_substructure}%
  \BibitemOpen
  \bibfield  {author} {\bibinfo {author} {\bibfnamefont {J.}~\bibnamefont
  {Brehmer}}, \bibinfo {author} {\bibfnamefont {S.}~\bibnamefont
  {Mishra-Sharma}}, \bibinfo {author} {\bibfnamefont {J.}~\bibnamefont
  {Hermans}}, \bibinfo {author} {\bibfnamefont {G.}~\bibnamefont {Louppe}}, \
  and\ \bibinfo {author} {\bibfnamefont {K.}~\bibnamefont {Cranmer}},\
  }\href@noop {} {\  (\bibinfo {year} {2019})},\ \Eprint
  {http://arxiv.org/abs/1909.02005} {arXiv:1909.02005 [astro-ph.CO]}
  \BibitemShut {NoStop}%
%%CITATION = ARXIV:1909.02005;%%
\bibitem [{\citenamefont {{Vegetti}}\ \emph {et~al.}(2014)\citenamefont
  {{Vegetti}}, \citenamefont {{Koopmans}}, \citenamefont {{Auger}},
  \citenamefont {{Treu}},\ and\ \citenamefont
  {{Bolton}}}]{2014MNRAS.442.2017V}%
  \BibitemOpen
  \bibfield  {author} {\bibinfo {author} {\bibfnamefont {S.}~\bibnamefont
  {{Vegetti}}}, \bibinfo {author} {\bibfnamefont {L.~V.~E.}\ \bibnamefont
  {{Koopmans}}}, \bibinfo {author} {\bibfnamefont {M.~W.}\ \bibnamefont
  {{Auger}}}, \bibinfo {author} {\bibfnamefont {T.}~\bibnamefont {{Treu}}}, \
  and\ \bibinfo {author} {\bibfnamefont {A.~S.}\ \bibnamefont {{Bolton}}},\
  }\href {\doibase 10.1093/mnras/stu943} {\bibfield  {journal} {\bibinfo
  {journal} {\mnras}\ }\textbf {\bibinfo {volume} {442}},\ \bibinfo {pages}
  {2017} (\bibinfo {year} {2014})},\ \Eprint {http://arxiv.org/abs/1405.3666}
  {arXiv:1405.3666 [astro-ph.GA]} \BibitemShut {NoStop}%
\bibitem [{\citenamefont {Ritondale}\ \emph {et~al.}(2019)\citenamefont
  {Ritondale}, \citenamefont {Vegetti}, \citenamefont {Despali}, \citenamefont
  {Auger}, \citenamefont {Koopmans},\ and\ \citenamefont
  {McKean}}]{bells_2018}%
  \BibitemOpen
  \bibfield  {author} {\bibinfo {author} {\bibfnamefont {E.}~\bibnamefont
  {Ritondale}}, \bibinfo {author} {\bibfnamefont {S.}~\bibnamefont {Vegetti}},
  \bibinfo {author} {\bibfnamefont {G.}~\bibnamefont {Despali}}, \bibinfo
  {author} {\bibfnamefont {M.~W.}\ \bibnamefont {Auger}}, \bibinfo {author}
  {\bibfnamefont {L.~V.~E.}\ \bibnamefont {Koopmans}}, \ and\ \bibinfo {author}
  {\bibfnamefont {J.~P.}\ \bibnamefont {McKean}},\ }\href {\doibase
  10.1093/mnras/stz464} {\bibfield  {journal} {\bibinfo  {journal} {Mon. Not.
  Roy. Astron. Soc.}\ }\textbf {\bibinfo {volume} {485}},\ \bibinfo {pages}
  {2179} (\bibinfo {year} {2019})},\ \Eprint {http://arxiv.org/abs/1811.03627}
  {arXiv:1811.03627 [astro-ph.CO]} \BibitemShut {NoStop}%
%%CITATION = ARXIV:1811.03627;%%
\bibitem [{\citenamefont {{Oguri}}\ and\ \citenamefont
  {{Marshall}}(2010)}]{LSST_lenses}%
  \BibitemOpen
  \bibfield  {author} {\bibinfo {author} {\bibfnamefont {M.}~\bibnamefont
  {{Oguri}}}\ and\ \bibinfo {author} {\bibfnamefont {P.~J.}\ \bibnamefont
  {{Marshall}}},\ }\href {\doibase 10.1111/j.1365-2966.2010.16639.x} {\bibfield
   {journal} {\bibinfo  {journal} {\mnras}\ }\textbf {\bibinfo {volume}
  {405}},\ \bibinfo {pages} {2579} (\bibinfo {year} {2010})},\ \Eprint
  {http://arxiv.org/abs/1001.2037} {arXiv:1001.2037 [astro-ph.CO]} \BibitemShut
  {NoStop}%
\bibitem [{\citenamefont {{Pawase}}\ \emph {et~al.}(2014)\citenamefont
  {{Pawase}}, \citenamefont {{Courbin}}, \citenamefont {{Faure}}, \citenamefont
  {{Kokotanekova}},\ and\ \citenamefont {{Meylan}}}]{2014MNRAS.439.3392P}%
  \BibitemOpen
  \bibfield  {author} {\bibinfo {author} {\bibfnamefont {R.~S.}\ \bibnamefont
  {{Pawase}}}, \bibinfo {author} {\bibfnamefont {F.}~\bibnamefont {{Courbin}}},
  \bibinfo {author} {\bibfnamefont {C.}~\bibnamefont {{Faure}}}, \bibinfo
  {author} {\bibfnamefont {R.}~\bibnamefont {{Kokotanekova}}}, \ and\ \bibinfo
  {author} {\bibfnamefont {G.}~\bibnamefont {{Meylan}}},\ }\href {\doibase
  10.1093/mnras/stu179} {\bibfield  {journal} {\bibinfo  {journal} {\mnras}\
  }\textbf {\bibinfo {volume} {439}},\ \bibinfo {pages} {3392} (\bibinfo {year}
  {2014})},\ \Eprint {http://arxiv.org/abs/1206.3412} {arXiv:1206.3412
  [astro-ph.CO]} \BibitemShut {NoStop}%
\bibitem [{\citenamefont {Collett}(2015)}]{Collett_2015}%
  \BibitemOpen
  \bibfield  {author} {\bibinfo {author} {\bibfnamefont {T.~E.}\ \bibnamefont
  {Collett}},\ }\href {\doibase 10.1088/0004-637x/811/1/20} {\bibfield
  {journal} {\bibinfo  {journal} {The Astrophysical Journal}\ }\textbf
  {\bibinfo {volume} {811}},\ \bibinfo {pages} {20} (\bibinfo {year}
  {2015})}\BibitemShut {NoStop}%
\bibitem [{\citenamefont {Lecun}\ \emph {et~al.}(1998)\citenamefont {Lecun},
  \citenamefont {Bottou}, \citenamefont {Bengio},\ and\ \citenamefont
  {Haffner}}]{cnn}%
  \BibitemOpen
  \bibfield  {author} {\bibinfo {author} {\bibfnamefont {Y.}~\bibnamefont
  {Lecun}}, \bibinfo {author} {\bibfnamefont {L.}~\bibnamefont {Bottou}},
  \bibinfo {author} {\bibfnamefont {Y.}~\bibnamefont {Bengio}}, \ and\ \bibinfo
  {author} {\bibfnamefont {P.}~\bibnamefont {Haffner}},\ }in\ \href@noop {}
  {\emph {\bibinfo {booktitle} {Proceedings of the IEEE}}}\ (\bibinfo {year}
  {1998})\ pp.\ \bibinfo {pages} {2278--2324}\BibitemShut {NoStop}%
\bibitem [{\citenamefont {LeCun}\ and\ \citenamefont {Cortes}(2010)}]{mnist}%
  \BibitemOpen
  \bibfield  {author} {\bibinfo {author} {\bibfnamefont {Y.}~\bibnamefont
  {LeCun}}\ and\ \bibinfo {author} {\bibfnamefont {C.}~\bibnamefont {Cortes}},\
  }\href {http://yann.lecun.com/exdb/mnist/} {\  (\bibinfo {year}
  {2010})}\BibitemShut {NoStop}%
\bibitem [{\citenamefont {Krizhevsky}(2009)}]{cifar10}%
  \BibitemOpen
  \bibfield  {author} {\bibinfo {author} {\bibfnamefont {A.}~\bibnamefont
  {Krizhevsky}},\ }\href@noop {} {\emph {\bibinfo {title} {Learning multiple
  layers of features from tiny images}}},\ \bibinfo {type} {Tech. Rep.}\
  (\bibinfo {year} {2009})\BibitemShut {NoStop}%
\bibitem [{\citenamefont {{Schaefer}}\ \emph {et~al.}(2018)\citenamefont
  {{Schaefer}}, \citenamefont {{Geiger}}, \citenamefont {{Kuntzer}},\ and\
  \citenamefont {{Kneib}}}]{2018A&A...611A...2S}%
  \BibitemOpen
  \bibfield  {author} {\bibinfo {author} {\bibfnamefont {C.}~\bibnamefont
  {{Schaefer}}}, \bibinfo {author} {\bibfnamefont {M.}~\bibnamefont
  {{Geiger}}}, \bibinfo {author} {\bibfnamefont {T.}~\bibnamefont {{Kuntzer}}},
  \ and\ \bibinfo {author} {\bibfnamefont {J.~P.}\ \bibnamefont {{Kneib}}},\
  }\href {\doibase 10.1051/0004-6361/201731201} {\bibfield  {journal} {\bibinfo
   {journal} {\aap}\ }\textbf {\bibinfo {volume} {611}},\ \bibinfo {eid} {A2}
  (\bibinfo {year} {2018})},\ \Eprint {http://arxiv.org/abs/1705.07132}
  {arXiv:1705.07132 [astro-ph.IM]} \BibitemShut {NoStop}%
\bibitem [{\citenamefont {{Davies}}\ \emph {et~al.}(2019)\citenamefont
  {{Davies}}, \citenamefont {{Serjeant}},\ and\ \citenamefont
  {{Bromley}}}]{2019MNRAS.487.5263D}%
  \BibitemOpen
  \bibfield  {author} {\bibinfo {author} {\bibfnamefont {A.}~\bibnamefont
  {{Davies}}}, \bibinfo {author} {\bibfnamefont {S.}~\bibnamefont
  {{Serjeant}}}, \ and\ \bibinfo {author} {\bibfnamefont {J.~M.}\ \bibnamefont
  {{Bromley}}},\ }\href {\doibase 10.1093/mnras/stz1288} {\bibfield  {journal}
  {\bibinfo  {journal} {\mnras}\ }\textbf {\bibinfo {volume} {487}},\ \bibinfo
  {pages} {5263} (\bibinfo {year} {2019})},\ \Eprint
  {http://arxiv.org/abs/1905.04303} {arXiv:1905.04303 [astro-ph.IM]}
  \BibitemShut {NoStop}%
\bibitem [{\citenamefont {Hezaveh}\ \emph {et~al.}(2017)\citenamefont
  {Hezaveh}, \citenamefont {Perreault~Levasseur},\ and\ \citenamefont
  {Marshall}}]{Hezaveh_cnn1}%
  \BibitemOpen
  \bibfield  {author} {\bibinfo {author} {\bibfnamefont {Y.~D.}\ \bibnamefont
  {Hezaveh}}, \bibinfo {author} {\bibfnamefont {L.}~\bibnamefont
  {Perreault~Levasseur}}, \ and\ \bibinfo {author} {\bibfnamefont {P.~J.}\
  \bibnamefont {Marshall}},\ }\href {\doibase 10.1038/nature23463} {\bibfield
  {journal} {\bibinfo  {journal} {Nature}\ }\textbf {\bibinfo {volume} {548}},\
  \bibinfo {pages} {555} (\bibinfo {year} {2017})},\ \Eprint
  {http://arxiv.org/abs/1708.08842} {arXiv:1708.08842 [astro-ph.IM]}
  \BibitemShut {NoStop}%
%%CITATION = ARXIV:1708.08842;%%
\bibitem [{\citenamefont {Perreault~Levasseur}\ \emph
  {et~al.}(2017)\citenamefont {Perreault~Levasseur}, \citenamefont {Hezaveh},\
  and\ \citenamefont {Wechsler}}]{PerreaultLevasseur_cnn2}%
  \BibitemOpen
  \bibfield  {author} {\bibinfo {author} {\bibfnamefont {L.}~\bibnamefont
  {Perreault~Levasseur}}, \bibinfo {author} {\bibfnamefont {Y.~D.}\
  \bibnamefont {Hezaveh}}, \ and\ \bibinfo {author} {\bibfnamefont {R.~H.}\
  \bibnamefont {Wechsler}},\ }\href {\doibase 10.3847/2041-8213/aa9704}
  {\bibfield  {journal} {\bibinfo  {journal} {Astrophys. J.}\ }\textbf
  {\bibinfo {volume} {850}},\ \bibinfo {pages} {L7} (\bibinfo {year} {2017})},\
  \Eprint {http://arxiv.org/abs/1708.08843} {arXiv:1708.08843 [astro-ph.CO]}
  \BibitemShut {NoStop}%
%%CITATION = ARXIV:1708.08843;%%
\bibitem [{\citenamefont {Morningstar}\ \emph {et~al.}(2018)\citenamefont
  {Morningstar}, \citenamefont {Hezaveh}, \citenamefont {Perreault~Levasseur},
  \citenamefont {Blandford}, \citenamefont {Marshall}, \citenamefont {Putzky},\
  and\ \citenamefont {Wechsler}}]{Morningstar:2018ase}%
  \BibitemOpen
  \bibfield  {author} {\bibinfo {author} {\bibfnamefont {W.~R.}\ \bibnamefont
  {Morningstar}}, \bibinfo {author} {\bibfnamefont {Y.~D.}\ \bibnamefont
  {Hezaveh}}, \bibinfo {author} {\bibfnamefont {L.}~\bibnamefont
  {Perreault~Levasseur}}, \bibinfo {author} {\bibfnamefont {R.~D.}\
  \bibnamefont {Blandford}}, \bibinfo {author} {\bibfnamefont {P.~J.}\
  \bibnamefont {Marshall}}, \bibinfo {author} {\bibfnamefont {P.}~\bibnamefont
  {Putzky}}, \ and\ \bibinfo {author} {\bibfnamefont {R.~H.}\ \bibnamefont
  {Wechsler}},\ }\href@noop {} {\  (\bibinfo {year} {2018})},\ \Eprint
  {http://arxiv.org/abs/1808.00011} {arXiv:1808.00011 [astro-ph.IM]}
  \BibitemShut {NoStop}%
%%CITATION = ARXIV:1808.00011;%%
\bibitem [{\citenamefont {{Vegetti}}\ \emph {et~al.}(2010)\citenamefont
  {{Vegetti}}, \citenamefont {{Koopmans}}, \citenamefont {{Bolton}},
  \citenamefont {{Treu}},\ and\ \citenamefont
  {{Gavazzi}}}]{detection_2010_mnras}%
  \BibitemOpen
  \bibfield  {author} {\bibinfo {author} {\bibfnamefont {S.}~\bibnamefont
  {{Vegetti}}}, \bibinfo {author} {\bibfnamefont {L.~V.~E.}\ \bibnamefont
  {{Koopmans}}}, \bibinfo {author} {\bibfnamefont {A.}~\bibnamefont
  {{Bolton}}}, \bibinfo {author} {\bibfnamefont {T.}~\bibnamefont {{Treu}}}, \
  and\ \bibinfo {author} {\bibfnamefont {R.}~\bibnamefont {{Gavazzi}}},\ }\href
  {\doibase 10.1111/j.1365-2966.2010.16865.x} {\bibfield  {journal} {\bibinfo
  {journal} {\mnras}\ }\textbf {\bibinfo {volume} {408}},\ \bibinfo {pages}
  {1969} (\bibinfo {year} {2010})},\ \Eprint {http://arxiv.org/abs/0910.0760}
  {arXiv:0910.0760 [astro-ph.CO]} \BibitemShut {NoStop}%
\bibitem [{\citenamefont {{Vegetti}}\ \emph {et~al.}(2012)\citenamefont
  {{Vegetti}}, \citenamefont {{Lagattuta}}, \citenamefont {{McKean}},
  \citenamefont {{Auger}}, \citenamefont {{Fassnacht}},\ and\ \citenamefont
  {{Koopmans}}}]{vegetti_nature}%
  \BibitemOpen
  \bibfield  {author} {\bibinfo {author} {\bibfnamefont {S.}~\bibnamefont
  {{Vegetti}}}, \bibinfo {author} {\bibfnamefont {D.~J.}\ \bibnamefont
  {{Lagattuta}}}, \bibinfo {author} {\bibfnamefont {J.~P.}\ \bibnamefont
  {{McKean}}}, \bibinfo {author} {\bibfnamefont {M.~W.}\ \bibnamefont
  {{Auger}}}, \bibinfo {author} {\bibfnamefont {C.~D.}\ \bibnamefont
  {{Fassnacht}}}, \ and\ \bibinfo {author} {\bibfnamefont {L.~V.~E.}\
  \bibnamefont {{Koopmans}}},\ }\href {\doibase 10.1038/nature10669} {\bibfield
   {journal} {\bibinfo  {journal} {\nat}\ }\textbf {\bibinfo {volume} {481}},\
  \bibinfo {pages} {341} (\bibinfo {year} {2012})},\ \Eprint
  {http://arxiv.org/abs/1201.3643} {arXiv:1201.3643 [astro-ph.CO]} \BibitemShut
  {NoStop}%
\bibitem [{\citenamefont {Birrer}\ and\ \citenamefont
  {Amara}(2018)}]{lenstronomy}%
  \BibitemOpen
  \bibfield  {author} {\bibinfo {author} {\bibfnamefont {S.}~\bibnamefont
  {Birrer}}\ and\ \bibinfo {author} {\bibfnamefont {A.}~\bibnamefont {Amara}},\
  }\href {\doibase 10.1016/j.dark.2018.11.002} {\  (\bibinfo {year} {2018}),\
  10.1016/j.dark.2018.11.002},\ \Eprint {http://arxiv.org/abs/1803.09746}
  {arXiv:1803.09746 [astro-ph.CO]} \BibitemShut {NoStop}%
%%CITATION = ARXIV:1803.09746;%%
\bibitem [{\citenamefont {{Kormann}}\ \emph {et~al.}(1994)\citenamefont
  {{Kormann}}, \citenamefont {{Schneider}},\ and\ \citenamefont
  {{Bartelmann}}}]{SIE}%
  \BibitemOpen
  \bibfield  {author} {\bibinfo {author} {\bibfnamefont {R.}~\bibnamefont
  {{Kormann}}}, \bibinfo {author} {\bibfnamefont {P.}~\bibnamefont
  {{Schneider}}}, \ and\ \bibinfo {author} {\bibfnamefont {M.}~\bibnamefont
  {{Bartelmann}}},\ }\href@noop {} {\bibfield  {journal} {\bibinfo  {journal}
  {\aap}\ }\textbf {\bibinfo {volume} {284}},\ \bibinfo {pages} {285} (\bibinfo
  {year} {1994})}\BibitemShut {NoStop}%
\bibitem [{\citenamefont {Vogelsberger}\ \emph {et~al.}(2016)\citenamefont
  {Vogelsberger}, \citenamefont {Zavala}, \citenamefont {Cyr-Racine},
  \citenamefont {Pfrommer}, \citenamefont {Bringmann},\ and\ \citenamefont
  {Sigurdson}}]{ETHOS}%
  \BibitemOpen
  \bibfield  {author} {\bibinfo {author} {\bibfnamefont {M.}~\bibnamefont
  {Vogelsberger}}, \bibinfo {author} {\bibfnamefont {J.}~\bibnamefont
  {Zavala}}, \bibinfo {author} {\bibfnamefont {F.-Y.}\ \bibnamefont
  {Cyr-Racine}}, \bibinfo {author} {\bibfnamefont {C.}~\bibnamefont
  {Pfrommer}}, \bibinfo {author} {\bibfnamefont {T.}~\bibnamefont {Bringmann}},
  \ and\ \bibinfo {author} {\bibfnamefont {K.}~\bibnamefont {Sigurdson}},\
  }\href {\doibase 10.1093/mnras/stw1076} {\bibfield  {journal} {\bibinfo
  {journal} {Mon. Not. Roy. Astron. Soc.}\ }\textbf {\bibinfo {volume} {460}},\
  \bibinfo {pages} {1399} (\bibinfo {year} {2016})},\ \Eprint
  {http://arxiv.org/abs/1512.05349} {arXiv:1512.05349 [astro-ph.CO]}
  \BibitemShut {NoStop}%
%%CITATION = ARXIV:1512.05349;%%
\bibitem [{\citenamefont {Madau}\ \emph {et~al.}(2008)\citenamefont {Madau},
  \citenamefont {Diemand},\ and\ \citenamefont {Kuhlen}}]{Madau_2008}%
  \BibitemOpen
  \bibfield  {author} {\bibinfo {author} {\bibfnamefont {P.}~\bibnamefont
  {Madau}}, \bibinfo {author} {\bibfnamefont {J.}~\bibnamefont {Diemand}}, \
  and\ \bibinfo {author} {\bibfnamefont {M.}~\bibnamefont {Kuhlen}},\ }\href
  {\doibase 10.1086/587545} {\bibfield  {journal} {\bibinfo  {journal} {The
  Astrophysical Journal}\ }\textbf {\bibinfo {volume} {679}},\ \bibinfo {pages}
  {1260} (\bibinfo {year} {2008})}\BibitemShut {NoStop}%
\bibitem [{\citenamefont {Xu}\ \emph {et~al.}(2013)\citenamefont {Xu},
  \citenamefont {Sluse}, \citenamefont {Gao}, \citenamefont {Wang},
  \citenamefont {Frenk}, \citenamefont {Mao},\ and\ \citenamefont
  {Schneider}}]{Xu:2013kna}%
  \BibitemOpen
  \bibfield  {author} {\bibinfo {author} {\bibfnamefont {D.~D.}\ \bibnamefont
  {Xu}}, \bibinfo {author} {\bibfnamefont {D.}~\bibnamefont {Sluse}}, \bibinfo
  {author} {\bibfnamefont {L.}~\bibnamefont {Gao}}, \bibinfo {author}
  {\bibfnamefont {J.}~\bibnamefont {Wang}}, \bibinfo {author} {\bibfnamefont
  {C.}~\bibnamefont {Frenk}}, \bibinfo {author} {\bibfnamefont
  {S.}~\bibnamefont {Mao}}, \ and\ \bibinfo {author} {\bibfnamefont
  {P.}~\bibnamefont {Schneider}}\ }(\bibinfo {year} {2013})\ \Eprint
  {http://arxiv.org/abs/1307.4220} {arXiv:1307.4220 [astro-ph.CO]} \BibitemShut
  {NoStop}%
%%CITATION = ARXIV:1307.4220;%%
\bibitem [{\citenamefont {Diemand}\ \emph
  {et~al.}(2007{\natexlab{a}})\citenamefont {Diemand}, \citenamefont {Kuhlen},\
  and\ \citenamefont {Madau}}]{vialactea}%
  \BibitemOpen
  \bibfield  {author} {\bibinfo {author} {\bibfnamefont {J.}~\bibnamefont
  {Diemand}}, \bibinfo {author} {\bibfnamefont {M.}~\bibnamefont {Kuhlen}}, \
  and\ \bibinfo {author} {\bibfnamefont {P.}~\bibnamefont {Madau}},\ }\href
  {\doibase 10.1086/510736} {\bibfield  {journal} {\bibinfo  {journal}
  {Astrophys. J.}\ }\textbf {\bibinfo {volume} {657}},\ \bibinfo {pages} {262}
  (\bibinfo {year} {2007}{\natexlab{a}})},\ \Eprint
  {http://arxiv.org/abs/astro-ph/0611370} {arXiv:astro-ph/0611370 [astro-ph]}
  \BibitemShut {NoStop}%
%%CITATION = ASTRO-PH/0611370;%%
\bibitem [{\citenamefont {{Vegetti}}\ and\ \citenamefont
  {{Koopmans}}(2009)}]{2009MNRAS.400.1583V}%
  \BibitemOpen
  \bibfield  {author} {\bibinfo {author} {\bibfnamefont {S.}~\bibnamefont
  {{Vegetti}}}\ and\ \bibinfo {author} {\bibfnamefont {L.~V.~E.}\ \bibnamefont
  {{Koopmans}}},\ }\href {\doibase 10.1111/j.1365-2966.2009.15559.x} {\bibfield
   {journal} {\bibinfo  {journal} {\mnras}\ }\textbf {\bibinfo {volume}
  {400}},\ \bibinfo {pages} {1583} (\bibinfo {year} {2009})},\ \Eprint
  {http://arxiv.org/abs/0903.4752} {arXiv:0903.4752 [astro-ph.CO]} \BibitemShut
  {NoStop}%
\bibitem [{\citenamefont {Diemand}\ \emph
  {et~al.}(2007{\natexlab{b}})\citenamefont {Diemand}, \citenamefont {Kuhlen},\
  and\ \citenamefont {Madau}}]{Diemand:2006ik}%
  \BibitemOpen
  \bibfield  {author} {\bibinfo {author} {\bibfnamefont {J.}~\bibnamefont
  {Diemand}}, \bibinfo {author} {\bibfnamefont {M.}~\bibnamefont {Kuhlen}}, \
  and\ \bibinfo {author} {\bibfnamefont {P.}~\bibnamefont {Madau}},\ }\href
  {\doibase 10.1086/510736} {\bibfield  {journal} {\bibinfo  {journal}
  {Astrophys. J.}\ }\textbf {\bibinfo {volume} {657}},\ \bibinfo {pages} {262}
  (\bibinfo {year} {2007}{\natexlab{b}})},\ \Eprint
  {http://arxiv.org/abs/astro-ph/0611370} {arXiv:astro-ph/0611370 [astro-ph]}
  \BibitemShut {NoStop}%
%%CITATION = ASTRO-PH/0611370;%%
\bibitem [{\citenamefont {Diemand}\ \emph
  {et~al.}(2007{\natexlab{c}})\citenamefont {Diemand}, \citenamefont {Kuhlen},\
  and\ \citenamefont {Madau}}]{Diemand:2007qr}%
  \BibitemOpen
  \bibfield  {author} {\bibinfo {author} {\bibfnamefont {J.}~\bibnamefont
  {Diemand}}, \bibinfo {author} {\bibfnamefont {M.}~\bibnamefont {Kuhlen}}, \
  and\ \bibinfo {author} {\bibfnamefont {P.}~\bibnamefont {Madau}},\ }\href
  {\doibase 10.1086/520573} {\bibfield  {journal} {\bibinfo  {journal}
  {Astrophys. J.}\ }\textbf {\bibinfo {volume} {667}},\ \bibinfo {pages} {859}
  (\bibinfo {year} {2007}{\natexlab{c}})},\ \Eprint
  {http://arxiv.org/abs/astro-ph/0703337} {arXiv:astro-ph/0703337 [astro-ph]}
  \BibitemShut {NoStop}%
%%CITATION = ASTRO-PH/0703337;%%
\bibitem [{\citenamefont {Diemand}\ \emph {et~al.}(2008)\citenamefont
  {Diemand}, \citenamefont {Kuhlen}, \citenamefont {Madau}, \citenamefont
  {Zemp}, \citenamefont {Moore}, \citenamefont {Potter},\ and\ \citenamefont
  {Stadel}}]{Diemand:2008in}%
  \BibitemOpen
  \bibfield  {author} {\bibinfo {author} {\bibfnamefont {J.}~\bibnamefont
  {Diemand}}, \bibinfo {author} {\bibfnamefont {M.}~\bibnamefont {Kuhlen}},
  \bibinfo {author} {\bibfnamefont {P.}~\bibnamefont {Madau}}, \bibinfo
  {author} {\bibfnamefont {M.}~\bibnamefont {Zemp}}, \bibinfo {author}
  {\bibfnamefont {B.}~\bibnamefont {Moore}}, \bibinfo {author} {\bibfnamefont
  {D.}~\bibnamefont {Potter}}, \ and\ \bibinfo {author} {\bibfnamefont
  {J.}~\bibnamefont {Stadel}},\ }\href {\doibase 10.1038/nature07153}
  {\bibfield  {journal} {\bibinfo  {journal} {Nature}\ }\textbf {\bibinfo
  {volume} {454}},\ \bibinfo {pages} {735} (\bibinfo {year} {2008})},\ \Eprint
  {http://arxiv.org/abs/0805.1244} {arXiv:0805.1244 [astro-ph]} \BibitemShut
  {NoStop}%
%%CITATION = ARXIV:0805.1244;%%
\bibitem [{\citenamefont {Vegetti}\ \emph {et~al.}(2018)\citenamefont
  {Vegetti}, \citenamefont {Despali}, \citenamefont {Lovell},\ and\
  \citenamefont {Enzi}}]{Vegetti:2018dly}%
  \BibitemOpen
  \bibfield  {author} {\bibinfo {author} {\bibfnamefont {S.}~\bibnamefont
  {Vegetti}}, \bibinfo {author} {\bibfnamefont {G.}~\bibnamefont {Despali}},
  \bibinfo {author} {\bibfnamefont {M.~R.}\ \bibnamefont {Lovell}}, \ and\
  \bibinfo {author} {\bibfnamefont {W.}~\bibnamefont {Enzi}},\ }\href {\doibase
  10.1093/mnras/sty2393} {\bibfield  {journal} {\bibinfo  {journal} {Mon. Not.
  Roy. Astron. Soc.}\ }\textbf {\bibinfo {volume} {481}},\ \bibinfo {pages}
  {3661} (\bibinfo {year} {2018})},\ \Eprint {http://arxiv.org/abs/1801.01505}
  {arXiv:1801.01505 [astro-ph.CO]} \BibitemShut {NoStop}%
%%CITATION = ARXIV:1801.01505;%%
\bibitem [{\citenamefont {{Navarro}}\ \emph {et~al.}(1996)\citenamefont
  {{Navarro}}, \citenamefont {{Frenk}},\ and\ \citenamefont {{White}}}]{nfw}%
  \BibitemOpen
  \bibfield  {author} {\bibinfo {author} {\bibfnamefont {J.~F.}\ \bibnamefont
  {{Navarro}}}, \bibinfo {author} {\bibfnamefont {C.~S.}\ \bibnamefont
  {{Frenk}}}, \ and\ \bibinfo {author} {\bibfnamefont {S.~D.~M.}\ \bibnamefont
  {{White}}},\ }\href {\doibase 10.1086/177173} {\bibfield  {journal} {\bibinfo
   {journal} {\apj}\ }\textbf {\bibinfo {volume} {462}},\ \bibinfo {pages}
  {563} (\bibinfo {year} {1996})},\ \Eprint
  {http://arxiv.org/abs/astro-ph/9508025} {arXiv:astro-ph/9508025 [astro-ph]}
  \BibitemShut {NoStop}%
\bibitem [{\citenamefont {Ade}\ \emph {et~al.}(2016)\citenamefont {Ade} \emph
  {et~al.}}]{Planck2015}%
  \BibitemOpen
  \bibfield  {author} {\bibinfo {author} {\bibfnamefont {P.~A.~R.}\
  \bibnamefont {Ade}} \emph {et~al.} (\bibinfo {collaboration} {Planck}),\
  }\href {\doibase 10.1051/0004-6361/201525830} {\bibfield  {journal} {\bibinfo
   {journal} {Astron. Astrophys.}\ }\textbf {\bibinfo {volume} {594}},\
  \bibinfo {pages} {A13} (\bibinfo {year} {2016})},\ \Eprint
  {http://arxiv.org/abs/1502.01589} {arXiv:1502.01589 [astro-ph.CO]}
  \BibitemShut {NoStop}%
%%CITATION = ARXIV:1502.01589;%%
\bibitem [{\citenamefont {Springel}\ \emph {et~al.}(2008)\citenamefont
  {Springel}, \citenamefont {Wang}, \citenamefont {Vogelsberger}, \citenamefont
  {Ludlow}, \citenamefont {Jenkins}, \citenamefont {Helmi}, \citenamefont
  {Navarro}, \citenamefont {Frenk},\ and\ \citenamefont {White}}]{Aquarius}%
  \BibitemOpen
  \bibfield  {author} {\bibinfo {author} {\bibfnamefont {V.}~\bibnamefont
  {Springel}}, \bibinfo {author} {\bibfnamefont {J.}~\bibnamefont {Wang}},
  \bibinfo {author} {\bibfnamefont {M.}~\bibnamefont {Vogelsberger}}, \bibinfo
  {author} {\bibfnamefont {A.}~\bibnamefont {Ludlow}}, \bibinfo {author}
  {\bibfnamefont {A.}~\bibnamefont {Jenkins}}, \bibinfo {author} {\bibfnamefont
  {A.}~\bibnamefont {Helmi}}, \bibinfo {author} {\bibfnamefont {J.~F.}\
  \bibnamefont {Navarro}}, \bibinfo {author} {\bibfnamefont {C.~S.}\
  \bibnamefont {Frenk}}, \ and\ \bibinfo {author} {\bibfnamefont {S.~D.~M.}\
  \bibnamefont {White}},\ }\href {\doibase 10.1111/j.1365-2966.2008.14066.x}
  {\bibfield  {journal} {\bibinfo  {journal} {Mon. Not. Roy. Astron. Soc.}\
  }\textbf {\bibinfo {volume} {391}},\ \bibinfo {pages} {1685} (\bibinfo {year}
  {2008})},\ \Eprint {http://arxiv.org/abs/0809.0898} {arXiv:0809.0898
  [astro-ph]} \BibitemShut {NoStop}%
%%CITATION = ARXIV:0809.0898;%%
\bibitem [{\citenamefont {Cyr-Racine}\ \emph {et~al.}(2019)\citenamefont
  {Cyr-Racine}, \citenamefont {Keeton},\ and\ \citenamefont
  {Moustakas}}]{Cyr-Racine_2018}%
  \BibitemOpen
  \bibfield  {author} {\bibinfo {author} {\bibfnamefont {F.-Y.}\ \bibnamefont
  {Cyr-Racine}}, \bibinfo {author} {\bibfnamefont {C.~R.}\ \bibnamefont
  {Keeton}}, \ and\ \bibinfo {author} {\bibfnamefont {L.~A.}\ \bibnamefont
  {Moustakas}},\ }\href {\doibase 10.1103/PhysRevD.100.023013} {\bibfield
  {journal} {\bibinfo  {journal} {Phys. Rev.}\ }\textbf {\bibinfo {volume}
  {D100}},\ \bibinfo {pages} {023013} (\bibinfo {year} {2019})},\ \Eprint
  {http://arxiv.org/abs/1806.07897} {arXiv:1806.07897 [astro-ph.CO]}
  \BibitemShut {NoStop}%
%%CITATION = ARXIV:1806.07897;%%
\bibitem [{\citenamefont {{Fruchter}}\ and\ \citenamefont
  {{Hook}}(2002)}]{drizzling}%
  \BibitemOpen
  \bibfield  {author} {\bibinfo {author} {\bibfnamefont {A.~S.}\ \bibnamefont
  {{Fruchter}}}\ and\ \bibinfo {author} {\bibfnamefont {R.~N.}\ \bibnamefont
  {{Hook}}},\ }\href {\doibase 10.1086/338393} {\bibfield  {journal} {\bibinfo
  {journal} {\pasp}\ }\textbf {\bibinfo {volume} {114}},\ \bibinfo {pages}
  {144} (\bibinfo {year} {2002})},\ \Eprint
  {http://arxiv.org/abs/astro-ph/9808087} {arXiv:astro-ph/9808087 [astro-ph]}
  \BibitemShut {NoStop}%
\bibitem [{\citenamefont {{Kingma}}\ and\ \citenamefont
  {{Welling}}(2013)}]{reparametrization}%
  \BibitemOpen
  \bibfield  {author} {\bibinfo {author} {\bibfnamefont {D.~P.}\ \bibnamefont
  {{Kingma}}}\ and\ \bibinfo {author} {\bibfnamefont {M.}~\bibnamefont
  {{Welling}}},\ }\href@noop {} {\bibfield  {journal} {\bibinfo  {journal}
  {arXiv e-prints}\ ,\ \bibinfo {eid} {arXiv:1312.6114}} (\bibinfo {year}
  {2013})},\ \Eprint {http://arxiv.org/abs/1312.6114} {arXiv:1312.6114
  [stat.ML]} \BibitemShut {NoStop}%
\bibitem [{\citenamefont {Paszke}\ \emph {et~al.}(2017)\citenamefont {Paszke},
  \citenamefont {Gross}, \citenamefont {Chintala}, \citenamefont {Chanan},
  \citenamefont {Yang}, \citenamefont {DeVito}, \citenamefont {Lin},
  \citenamefont {Desmaison}, \citenamefont {Antiga},\ and\ \citenamefont
  {Lerer}}]{pytorch}%
  \BibitemOpen
  \bibfield  {author} {\bibinfo {author} {\bibfnamefont {A.}~\bibnamefont
  {Paszke}}, \bibinfo {author} {\bibfnamefont {S.}~\bibnamefont {Gross}},
  \bibinfo {author} {\bibfnamefont {S.}~\bibnamefont {Chintala}}, \bibinfo
  {author} {\bibfnamefont {G.}~\bibnamefont {Chanan}}, \bibinfo {author}
  {\bibfnamefont {E.}~\bibnamefont {Yang}}, \bibinfo {author} {\bibfnamefont
  {Z.}~\bibnamefont {DeVito}}, \bibinfo {author} {\bibfnamefont
  {Z.}~\bibnamefont {Lin}}, \bibinfo {author} {\bibfnamefont {A.}~\bibnamefont
  {Desmaison}}, \bibinfo {author} {\bibfnamefont {L.}~\bibnamefont {Antiga}}, \
  and\ \bibinfo {author} {\bibfnamefont {A.}~\bibnamefont {Lerer}},\ }in\
  \href@noop {} {\emph {\bibinfo {booktitle} {NIPS Autodiff Workshop}}}\
  (\bibinfo {year} {2017})\BibitemShut {NoStop}%
\bibitem [{\citenamefont {Glorot}\ and\ \citenamefont {Bengio}(2010)}]{xavier}%
  \BibitemOpen
  \bibfield  {author} {\bibinfo {author} {\bibfnamefont {X.}~\bibnamefont
  {Glorot}}\ and\ \bibinfo {author} {\bibfnamefont {Y.}~\bibnamefont
  {Bengio}},\ }in\ \href@noop {} {\emph {\bibinfo {booktitle} {In Proceedings
  of the International Conference on Artificial Intelligence and Statistics
  (AISTATS’10). Society for Artificial Intelligence and Statistics}}}\
  (\bibinfo {year} {2010})\BibitemShut {NoStop}%
\bibitem [{\citenamefont {Dalal}\ and\ \citenamefont
  {Kochanek}(2002)}]{Dalal:2001fq}%
  \BibitemOpen
  \bibfield  {author} {\bibinfo {author} {\bibfnamefont {N.}~\bibnamefont
  {Dalal}}\ and\ \bibinfo {author} {\bibfnamefont {C.~S.}\ \bibnamefont
  {Kochanek}},\ }\href {\doibase 10.1086/340303} {\bibfield  {journal}
  {\bibinfo  {journal} {Astrophys. J.}\ }\textbf {\bibinfo {volume} {572}},\
  \bibinfo {pages} {25} (\bibinfo {year} {2002})},\ \Eprint
  {http://arxiv.org/abs/astro-ph/0111456} {arXiv:astro-ph/0111456 [astro-ph]}
  \BibitemShut {NoStop}%
%%CITATION = ASTRO-PH/0111456;%%
\bibitem [{\citenamefont {{Minor}}\ \emph {et~al.}(2017)\citenamefont
  {{Minor}}, \citenamefont {{Kaplinghat}},\ and\ \citenamefont
  {{Li}}}]{effective_mass}%
  \BibitemOpen
  \bibfield  {author} {\bibinfo {author} {\bibfnamefont {Q.~E.}\ \bibnamefont
  {{Minor}}}, \bibinfo {author} {\bibfnamefont {M.}~\bibnamefont
  {{Kaplinghat}}}, \ and\ \bibinfo {author} {\bibfnamefont {N.}~\bibnamefont
  {{Li}}},\ }\href {\doibase 10.3847/1538-4357/aa7fee} {\bibfield  {journal}
  {\bibinfo  {journal} {\apj}\ }\textbf {\bibinfo {volume} {845}},\ \bibinfo
  {eid} {118} (\bibinfo {year} {2017})},\ \Eprint
  {http://arxiv.org/abs/1612.05250} {arXiv:1612.05250 [astro-ph.GA]}
  \BibitemShut {NoStop}%
\bibitem [{\citenamefont {Krizhevsky}\ \emph {et~al.}(2012)\citenamefont
  {Krizhevsky}, \citenamefont {Sutskever},\ and\ \citenamefont
  {Hinton}}]{imagenet}%
  \BibitemOpen
  \bibfield  {author} {\bibinfo {author} {\bibfnamefont {A.}~\bibnamefont
  {Krizhevsky}}, \bibinfo {author} {\bibfnamefont {I.}~\bibnamefont
  {Sutskever}}, \ and\ \bibinfo {author} {\bibfnamefont {G.~E.}\ \bibnamefont
  {Hinton}},\ }in\ \href
  {http://papers.nips.cc/paper/4824-imagenet-classification-with-deep-convolutional-neural-networks.pdf}
  {\emph {\bibinfo {booktitle} {Advances in Neural Information Processing
  Systems 25}}},\ \bibinfo {editor} {edited by\ \bibinfo {editor}
  {\bibfnamefont {F.}~\bibnamefont {Pereira}}, \bibinfo {editor} {\bibfnamefont
  {C.~J.~C.}\ \bibnamefont {Burges}}, \bibinfo {editor} {\bibfnamefont
  {L.}~\bibnamefont {Bottou}}, \ and\ \bibinfo {editor} {\bibfnamefont {K.~Q.}\
  \bibnamefont {Weinberger}}}\ (\bibinfo  {publisher} {Curran Associates,
  Inc.},\ \bibinfo {year} {2012})\ pp.\ \bibinfo {pages}
  {1097--1105}\BibitemShut {NoStop}%
\bibitem [{\citenamefont {{Szegedy}}\ \emph {et~al.}(2014)\citenamefont
  {{Szegedy}}, \citenamefont {{Liu}}, \citenamefont {{Jia}}, \citenamefont
  {{Sermanet}}, \citenamefont {{Reed}}, \citenamefont {{Anguelov}},
  \citenamefont {{Erhan}}, \citenamefont {{Vanhoucke}},\ and\ \citenamefont
  {{Rabinovich}}}]{googlenet}%
  \BibitemOpen
  \bibfield  {author} {\bibinfo {author} {\bibfnamefont {C.}~\bibnamefont
  {{Szegedy}}}, \bibinfo {author} {\bibfnamefont {W.}~\bibnamefont {{Liu}}},
  \bibinfo {author} {\bibfnamefont {Y.}~\bibnamefont {{Jia}}}, \bibinfo
  {author} {\bibfnamefont {P.}~\bibnamefont {{Sermanet}}}, \bibinfo {author}
  {\bibfnamefont {S.}~\bibnamefont {{Reed}}}, \bibinfo {author} {\bibfnamefont
  {D.}~\bibnamefont {{Anguelov}}}, \bibinfo {author} {\bibfnamefont
  {D.}~\bibnamefont {{Erhan}}}, \bibinfo {author} {\bibfnamefont
  {V.}~\bibnamefont {{Vanhoucke}}}, \ and\ \bibinfo {author} {\bibfnamefont
  {A.}~\bibnamefont {{Rabinovich}}},\ }\href@noop {} {\bibfield  {journal}
  {\bibinfo  {journal} {arXiv e-prints}\ ,\ \bibinfo {eid} {arXiv:1409.4842}}
  (\bibinfo {year} {2014})},\ \Eprint {http://arxiv.org/abs/1409.4842}
  {arXiv:1409.4842 [cs.CV]} \BibitemShut {NoStop}%
\bibitem [{\citenamefont {{He}}\ \emph {et~al.}(2015)\citenamefont {{He}},
  \citenamefont {{Zhang}}, \citenamefont {{Ren}},\ and\ \citenamefont
  {{Sun}}}]{resnet}%
  \BibitemOpen
  \bibfield  {author} {\bibinfo {author} {\bibfnamefont {K.}~\bibnamefont
  {{He}}}, \bibinfo {author} {\bibfnamefont {X.}~\bibnamefont {{Zhang}}},
  \bibinfo {author} {\bibfnamefont {S.}~\bibnamefont {{Ren}}}, \ and\ \bibinfo
  {author} {\bibfnamefont {J.}~\bibnamefont {{Sun}}},\ }\href@noop {}
  {\bibfield  {journal} {\bibinfo  {journal} {arXiv e-prints}\ ,\ \bibinfo
  {eid} {arXiv:1512.03385}} (\bibinfo {year} {2015})},\ \Eprint
  {http://arxiv.org/abs/1512.03385} {arXiv:1512.03385 [cs.CV]} \BibitemShut
  {NoStop}%
\bibitem [{\citenamefont {{Huang}}\ \emph {et~al.}(2016)\citenamefont
  {{Huang}}, \citenamefont {{Liu}}, \citenamefont {{van der Maaten}},\ and\
  \citenamefont {{Weinberger}}}]{densenet}%
  \BibitemOpen
  \bibfield  {author} {\bibinfo {author} {\bibfnamefont {G.}~\bibnamefont
  {{Huang}}}, \bibinfo {author} {\bibfnamefont {Z.}~\bibnamefont {{Liu}}},
  \bibinfo {author} {\bibfnamefont {L.}~\bibnamefont {{van der Maaten}}}, \
  and\ \bibinfo {author} {\bibfnamefont {K.~Q.}\ \bibnamefont {{Weinberger}}},\
  }\href@noop {} {\bibfield  {journal} {\bibinfo  {journal} {arXiv e-prints}\
  ,\ \bibinfo {eid} {arXiv:1608.06993}} (\bibinfo {year} {2016})},\ \Eprint
  {http://arxiv.org/abs/1608.06993} {arXiv:1608.06993 [cs.CV]} \BibitemShut
  {NoStop}%
\bibitem [{\citenamefont {Alexander}\ \emph {et~al.}(2019)\citenamefont
  {Alexander}, \citenamefont {Gleyzer}, \citenamefont {McDonough},
  \citenamefont {Toomey},\ and\ \citenamefont {Usai}}]{stephon}%
  \BibitemOpen
  \bibfield  {author} {\bibinfo {author} {\bibfnamefont {S.}~\bibnamefont
  {Alexander}}, \bibinfo {author} {\bibfnamefont {S.}~\bibnamefont {Gleyzer}},
  \bibinfo {author} {\bibfnamefont {E.}~\bibnamefont {McDonough}}, \bibinfo
  {author} {\bibfnamefont {M.~W.}\ \bibnamefont {Toomey}}, \ and\ \bibinfo
  {author} {\bibfnamefont {E.}~\bibnamefont {Usai}},\ }\href@noop {} {\
  (\bibinfo {year} {2019})},\ \Eprint {http://arxiv.org/abs/1909.07346}
  {arXiv:1909.07346 [astro-ph.CO]} \BibitemShut {NoStop}%
%%CITATION = ARXIV:1909.07346;%%
\bibitem [{\citenamefont {Li}\ \emph {et~al.}(2017)\citenamefont {Li},
  \citenamefont {Frenk}, \citenamefont {Cole}, \citenamefont {Wang},\ and\
  \citenamefont {Gao}}]{Li_los}%
  \BibitemOpen
  \bibfield  {author} {\bibinfo {author} {\bibfnamefont {R.}~\bibnamefont
  {Li}}, \bibinfo {author} {\bibfnamefont {C.~S.}\ \bibnamefont {Frenk}},
  \bibinfo {author} {\bibfnamefont {S.}~\bibnamefont {Cole}}, \bibinfo {author}
  {\bibfnamefont {Q.}~\bibnamefont {Wang}}, \ and\ \bibinfo {author}
  {\bibfnamefont {L.}~\bibnamefont {Gao}},\ }\href {\doibase
  10.1093/mnras/stx554} {\bibfield  {journal} {\bibinfo  {journal} {Mon. Not.
  Roy. Astron. Soc.}\ }\textbf {\bibinfo {volume} {468}},\ \bibinfo {pages}
  {1426} (\bibinfo {year} {2017})},\ \Eprint {http://arxiv.org/abs/1612.06227}
  {arXiv:1612.06227 [astro-ph.CO]} \BibitemShut {NoStop}%
%%CITATION = ARXIV:1612.06227;%%
\bibitem [{\citenamefont {Despali}\ \emph {et~al.}(2018)\citenamefont
  {Despali}, \citenamefont {Vegetti}, \citenamefont {White}, \citenamefont
  {Giocoli},\ and\ \citenamefont {van~den Bosch}}]{Despali_los}%
  \BibitemOpen
  \bibfield  {author} {\bibinfo {author} {\bibfnamefont {G.}~\bibnamefont
  {Despali}}, \bibinfo {author} {\bibfnamefont {S.}~\bibnamefont {Vegetti}},
  \bibinfo {author} {\bibfnamefont {S.~D.~M.}\ \bibnamefont {White}}, \bibinfo
  {author} {\bibfnamefont {C.}~\bibnamefont {Giocoli}}, \ and\ \bibinfo
  {author} {\bibfnamefont {F.~C.}\ \bibnamefont {van~den Bosch}},\ }\href
  {\doibase 10.1093/mnras/sty159} {\bibfield  {journal} {\bibinfo  {journal}
  {Mon. Not. Roy. Astron. Soc.}\ }\textbf {\bibinfo {volume} {475}},\ \bibinfo
  {pages} {5424} (\bibinfo {year} {2018})},\ \Eprint
  {http://arxiv.org/abs/1710.05029} {arXiv:1710.05029 [astro-ph.CO]}
  \BibitemShut {NoStop}%
%%CITATION = ARXIV:1710.05029;%%
\bibitem [{\citenamefont {Morningstar}\ \emph {et~al.}(2019)\citenamefont
  {Morningstar}, \citenamefont {Perreault~Levasseur}, \citenamefont {Hezaveh},
  \citenamefont {Blandford}, \citenamefont {Marshall}, \citenamefont {Putzky},
  \citenamefont {Rueter}, \citenamefont {Wechsler},\ and\ \citenamefont
  {Welling}}]{Morningstar:2019szx}%
  \BibitemOpen
  \bibfield  {author} {\bibinfo {author} {\bibfnamefont {W.~R.}\ \bibnamefont
  {Morningstar}}, \bibinfo {author} {\bibfnamefont {L.}~\bibnamefont
  {Perreault~Levasseur}}, \bibinfo {author} {\bibfnamefont {Y.~D.}\
  \bibnamefont {Hezaveh}}, \bibinfo {author} {\bibfnamefont {R.}~\bibnamefont
  {Blandford}}, \bibinfo {author} {\bibfnamefont {P.}~\bibnamefont {Marshall}},
  \bibinfo {author} {\bibfnamefont {P.}~\bibnamefont {Putzky}}, \bibinfo
  {author} {\bibfnamefont {T.~D.}\ \bibnamefont {Rueter}}, \bibinfo {author}
  {\bibfnamefont {R.}~\bibnamefont {Wechsler}}, \ and\ \bibinfo {author}
  {\bibfnamefont {M.}~\bibnamefont {Welling}},\ }\href@noop {} {\  (\bibinfo
  {year} {2019})},\ \Eprint {http://arxiv.org/abs/1901.01359} {arXiv:1901.01359
  [astro-ph.IM]} \BibitemShut {NoStop}%
%%CITATION = ARXIV:1901.01359;%%
\end{thebibliography}%

\end{document}